# Stochastic storage models in theoretical physics problems

V. V. Ryazanov

Institute for Nuclear Research, pr. Nauki, 47 Kiev, Ukraine, e-mail: vryazan19@gmail.com
ORCID id 0000-0002-5308-3212

Stochastic storage models based on essentially non-Gaussian noise are considered. The stochastic description of physical systems based on stochastic storage models is associated with generalized Poisson (or shot) noise, in which the jump values can be quite large. Stochastic storage models have a direct physical meaning: some elements enter the system and leave it. Storage processes fit into the general scheme of dynamic systems subject to the additive influence of a random process. The main relationships of storage models are described, and the possibilities of applying the mathematical provisions of stochastic storage processes to various physical problems are indicated. A number of examples of applying the stochastic storage model are considered.

## 1. Introduction

Stochastic models of storage (for example [5-9]) are one of the broad and ramified sections of stochastics. The generality of the mathematical theory allows it to be applied to arbitrary, not only physical problems. These models have a direct physical meaning: some elements enter the system and exit it. Models of the storage theory are related to queuing theory, risk theory, reliability theory, insurance theory, models of biological populations, sequential analysis, etc.

In this paper, we indicate the possibilities of applying the mathematical principles of stochastic storage processes to various physical problems: nonequilibrium thermodynamics [10-11], aerosol theory Refs. [12-17], general classification of the behavior of physical systems [18], probabilistic assessment of the safety of nuclear facilities [19-20], tree growth processes [21], Gibbs statistics [22], noise-induced phase transitions [4, 22-23], neutron processes in nuclear reactors [24-26], micelle formation [27], raft-like domains in biological cell membranes [28-30], etc.

One of the first works of storage theory is Ref. [31]. In physical problems, random walks in a random environment are used [32-33], which are also associated with storage models. In studies of storage models, such mathematical apparatus as a combination of various methods of the theory of Markov chains, Markov processes of diffusion type, regenerative processes, restoration processes, random walks, including in a random environment are used.

For the class of stochastic storage models Refs. [5-9], the following physical prerequisites can be distinguished: (i) restriction to a positive half-space of states, (ii) jumps in random physical quantities are not necessarily assumed to be small; (iii) a substantially non-zero thermodynamic flow specified by the random entry process.

The basic, typical probability distributions in storage models are not Gaussian, but rather exponential and gamma distributions, characteristic, for example, of the Tsallis statistics Ref. [34]. And in inventory management models, it was assumed that the demand size has a gamma distribution (e.g., Ref. [35]). Storage processes can describe states far from equilibrium. Storage models are applied to the results of nonlinear nonequilibrium thermodynamics Refs. [10, 11, 36].

Simple cases of storage models do not require special probabilistic techniques and can be compared with kinetic equations. Such approaches allow modeling the kinetics of open systems in which the entry rate is determined from physical considerations.

Within the framework of the stochastic storage model, relations for describing the behavior of thermodynamic systems are obtained. Explicit expressions are obtained for the kinetic potential that determines the nonequilibrium behavior of statistical systems, for its image, for flows and thermodynamic functions in a system with disturbances. A generalization of stochastic storage models is carried out and lifetimes (first- passage times) are considered. Knowing the patterns of lifetime behavior, one can set problems of searching for targeted effects for changing and controlling lifetimes.

For stochastic storage processes, it has been established that the finiteness of the system's lifetime is related to stationary conditions and corresponds to its complex macroscopic behavior, which depends on the characteristics of the system's interaction with the environment. In a system with an infinite lifetime, the average values behave unambiguously. This conclusion coincides with the conclusions of the Ref. [37] for the absence or presence of a thermodynamic limit.

The article is organized as follows. Sections 1–3 provide definitions of the storage process, introduce the kinetic potential and its image, flows, and consider the conditions for the stationary of storage processes. Section 4 introduces distributions containing thermodynamic forces. Averaging over this distribution makes it possible to compare the kinetic equations of the storage process with the phenomenological equations of motion. Section 5 examines macroscopic equations for storage models with stationary thermodynamic forces. Section 6 derives expressions for the exact equations of storage processes. Section 7 examines phase transitions. Section 8 introduces the quasi-potential and examines the influence of external forces on the evolution of the system. Section 9 provides an example of applying a storage model to describe the formation of micelles. Section 10 presents another approach to storage models, combining them with dynamic systems. Section 11 provides examples of applying storage processes to problems of tribology, aerosol systems, and raft kinetics in biological membranes. Brief conclusions are given in the conclusion.

## 1. Stochastic storage model

The stochastic storage model has a simple physical meaning. Some quantities randomly enter the system (often compared to a reservoir), forming a random value of the stock of these quantities in the system $X(t)$. The output from the system is described by a function depending on $X(t)$. There is also a close connection with the theory of dynamic systems with random effects (Sec. 10). The stochastic storage process is described by the equation Refs. [5-9]:

$$\frac{dX(t)}{dt}=\frac{dA}{dt}-r_{\chi}[X(t)], \qquad X(t)=X(0)+A(t)-\int_0^t r_{\chi}\left[X(u)\right]du, \tag{1}$$

where $X(t)$ is a random variable of the stock in the system, $dA(t)/dt$ is a random rate of stock entry into the system, $r_{\chi}[X(t)]$ is the exit rate. The exit rate takes into account the singularity at zero: $r_{\chi}[X(t)]=r(X(T))-r(0+)\chi_{X(t)}$; $\chi_X=1,\ at\ X=0,\ \ \chi_X=0,\ at\ X>0$. This is due to the fact that the storage model (1) is defined for non-negative values $X(t)\geq 0$, and the exit from an empty system must be zero. The function $r[X(t)]$ can be chosen arbitrarily. The entry function $A(t)$ can be described by various classes of random processes. In Refs. [5-7], the entry rate is characterized by a Levy process with non-decreasing trajectories and zero drift. Brownian motion and the Poisson process also belong to Levy processes. The Laplace transform $\boldsymbol{E}(\exp\{-\theta A(t)\})$ of the entry function $A(t)$ is associated with the function $\varphi(\theta)$, the so-called the scaled cumulant generating function (*SCGF*) or in shirt cumulant of a process $A(t)$ of the form

$$\boldsymbol{E}(\exp\{-\theta A(t)\} = \exp\{-t\varphi(\theta)\} = \int_0^\infty e^{-\theta x} k(x,t)dx\,; \qquad \varphi(\theta) = \int_0^\infty (1-\exp\{-\theta x\})\lambda g(x)dx\,, \quad (2)$$

$$\lambda = \varphi(\infty) < \infty; \quad \rho \equiv \lambda \int x g(x)dx = E(\frac{dA}{dt}) = \frac{d\varphi(\theta)}{d\theta}_{|\theta=0}; \quad \mu^{-1} \equiv \int x g(x)dx = \frac{1}{\varphi(\infty)} \frac{d\varphi(\theta)}{d\theta}_{|\theta=0},$$

where $\boldsymbol{E}(...)$ denotes averaging, $k(x,t) = P(x \le A(t) \le x + dx)$ is the probability density function of the random variable $A(t)$, $\lambda$ is the intensity of the input flow, and the density function $g(y)$ describes the magnitude of the jumps in the inflows. In representation (2), the input process corresponds to a generalized Poisson process Refs. [5, 9], where inflows occur at random moments in time in randomly sized portions. If we use the analogy with a reservoir and the water supply in it, then the quantity $\lambda$ describes the intensity of Poisson random jumps, moments in time when water inflows into the system occur. The density function $g(y) = P(y \le m \le y + dy)$ describes the random amount of water $m$ entering the reservoir in one jump, with an average of $\mu^{-1}$.

Any functions $B_\gamma(z)$ of dynamic variables $z=(p,q)=(q_1,...,q_n; p_1,...,p_n)$, sets of coordinates and momenta, having a macroscopic nature, are by definition random internal thermodynamic parameters Ref. [36]. These parameters can be energy, number of particles, etc. We assume that these include variables $X(t)$ from (1).

For the Laplace transform $\mathbf{E}(e^{-\theta B(t)})=F(e^{-\theta})=\int e^{-\theta y}\omega(y,t)dy$ of the probability distribution $\omega(B, t)$ of the internal thermodynamic parameter B of the storage model (1)-(2) with $X(t) = B(t)$ (X(t) – from (1)) we obtain an equation of the form Ref. [22]:

$$\frac{\partial F(e^{-\theta})}{\partial t} = [-\beta\varphi(\frac{\theta}{\beta}) + \theta r_{\chi_{-\partial/\partial\theta}}(-\frac{\partial}{\partial\theta})]F(e^{-\theta})\,, \qquad (3)$$

where $\beta^{-1}$ is a small parameter; for an equilibrium Gibbs system $\beta=1/k_BT_{eq}$, $k_B$ is the Boltzmann constant, $T_{eq}$ is the equilibrium temperature.

In the theory of random processes, the lifetime of a system $\Gamma$ (or first-passage time, (FPT)) is defined as the random time until the moment of the first achievement of a certain (zero) level by a random process $y(t)$, describing the behavior of the parameter $B(t)$, i.e.:

$$\Gamma_x = \inf\{t : y(t) = 0\}, \quad y(0) = x > 0\,. \qquad (4)$$

The characteristics of $\Gamma$ depend on the process $y(t)$. It can be shown Refs. [38, 39] that the distribution for the lifetime (4) obeys the Hermitian adjoint operator and for the Laplace transform $L(x,s)$ (5) of the distribution of the lifetime $P(\Gamma = y|x)$ of the system with the initial value $x$ of the process $y$, the moment of reaching which is considered,

$$\boldsymbol{E}(e^{-s\Gamma(x)}) \equiv L(x,s) = \int e^{-sy} p(\Gamma = y|x)dy\,; \qquad (5)$$

the equation conjugated to the equation (3) is written Refs. [22, 38, 39]:

$$[-\varphi(-\frac{\partial}{\partial x}) - r_\chi(x)\frac{\partial}{\partial x}]L(x,s) = sL(x,s) \qquad (6)$$

with boundary and initial conditions $L(0,s) = L(x,0) = 1$. The moments of the random lifetime are determined by differentiating the Laplace transform $L(x,s)$ (5) with respect to $s$.

For $F(e^{-\theta},s) = \int_0^\infty e^{-st}F(e^{-\theta},t)dt$ the equation (3), it is written as:

$$sF(e^{-\theta},s) - F(e^{-\theta},t=0) = [-\beta\varphi(\theta/\beta) + \theta r_{\chi_{-\partial/\partial\theta}}(-\partial/\partial\theta)]F(e^{-\theta},s)\,.$$

For an output function of the form:

$$r(q) = a(1-\chi_q) \qquad (7)$$

equation (3) takes the form:

$$\partial F_a(e^{-\theta},t)/\partial t) = [-\beta\varphi(\theta/\beta) + \theta r_{\chi_{-\frac{\partial}{\partial\theta}}}(-\partial/\partial\theta)]F_a(e^{-\theta},t) = [-\beta\varphi(\theta/\beta) + \theta a]F_a(e^{-\theta},t) - a\theta P_0(t)\,. \quad (8)$$

The fact that $\chi_{-\partial/\partial\theta} F(\exp\{-\theta\}, t)=P_0(t)$.

Output model:

$$r(q) = a(1-\chi_q) + bq\,, \quad \chi_q = 1,\; q = 0; \quad \chi_q = 0, \quad q > 0 \quad (9)$$

is a combination of model (7) and model (22). With an output function of the form (9), equation (3) takes the form:

$$\frac{\partial F(e^x)}{\partial t} = x[-a + \frac{\beta\varphi(-x/\beta)}{(-x)}]F(e^x,t) + axP_0(t) - bx\frac{\partial}{\partial x}F(e^x,t)\,. \quad (10)$$

In the stationary case, equation (10) has a solution:

$$F_{a+bq,st}(e^{-\theta}) = e^{-\int_0^\theta \frac{1}{b}[\frac{\beta\varphi(u/\beta)}{u}du - a\theta]}[1 - \frac{aP_0}{b}\int_0^\theta e^{\int_0^y \frac{1}{b}[\frac{\beta\varphi(u/\beta)}{u}du - ay]}dy]; \quad P_0^{-1} = \frac{a}{b}\int_0^\infty e^{\frac{1}{b}\int_0^y \frac{1}{b}[\frac{\beta\varphi(u/\beta)}{u}du - ay]}dy\,, \quad (11)$$

coinciding with the results of Ref. [7]. The solution of the equation for mean $\langle q(t)\rangle$, obtained from equation (10), has the form:

$$\langle q(t)\rangle = \langle n(t)\rangle = (q_0 + \frac{a-\lambda}{b})e^{-bt} - \frac{a-\lambda}{b} + \int_0^t e^{-(t-\tau)b}aP_0(\tau)d\tau, \quad q_0 = n(t=0)\,. \quad (12)$$

This expression (12) turns into a solution of equation (8) when $b\to 0$ and $r(x)\to a(1-\chi_x)$. The Laplace transform of expression (12) with respect to time is equal to:

$$\langle q(s)\rangle = [(q_0 + aP_0(s))s - (a-\rho)]/s(s+b)\,. \quad (13)$$

From where:

$$\langle q_{st}\rangle = \frac{\rho - a}{b} + \frac{aP_0}{b}; \quad P_0 = \frac{a-\rho+b\langle q_{st}\rangle}{a}\,. \quad (14)$$

We also get that:

$$\langle q^2(t)\rangle = (q_0^2 - \frac{\sigma^2}{2b})e^{-2bt} + \frac{\sigma^2}{2b} - 2(a-\rho)\int_0^t q(\tau)e^{-2(t-\tau)b}d\tau\,; \quad (15)$$

$$< q^2{}_{st} > = \sigma^2/2b + (\rho - a)(\rho - a + aP_0)/b^2\,. \quad (16)$$

By specifying explicit expressions for the density of the distribution function $g(x)$ of the magnitude of the jumps in receipts from (2), we can find the function $\varphi(s)$. In a homogeneous Poisson process, the distance between two successive jumps in the trajectory has an exponential distribution. In a wide variety of areas, the time intervals between the occurrences of events are described by an exponential distribution. Let us apply this distribution to the description of the magnitude of one jump. Let us consider a distribution of the form:

$$g(x) = \mu e^{-\mu x}\,, \quad \varphi(\theta) = \lambda\theta/(\mu+\theta)\,. \quad (17)$$

Real physical systems can be described by more complex functions, but the distribution (17) allows us to obtain explicit analytical expressions for the average $A(x)$, thermodynamic forces $x(A)$ from (44), the function $dA/dt = f(A)$ (45), etc. The possibilities of applying stochastic storage processes to various problems are described in Refs. [4, 22].

When choosing an exponential distribution of the form (17) in (2) and the output function (9), the stationary probability density corresponding to the Laplace transform (11) is equal to:

$$\omega_{st}(x) = P_0[\delta(x) + f_{st}(x)], \qquad \omega_{st}(x) = P_0[\delta(x) + (\lambda/b)a^{-\frac{\lambda}{b}}\exp\{-\mu x\}(a+bx)^{\frac{\lambda}{b}-1}]\,, \quad (18)$$

$$P_0^{-1} = 1 + (\mu a/b)^{-\frac{\lambda}{b}}(\lambda/b)\exp\{\mu a/b\}\Gamma(\lambda/b,\, \mu a/b) \ , \tag{19}$$

where $\Gamma(\lambda/b;\, \mu a/b)$ is the incomplete gamma function Ref. [40], $\delta(x)$ is the delta function, the Dirac function Ref. [40]. In (18)-(19) the feature of the distribution at zero, characterized by the delta function, and the presence of a continuous part of the stationary distribution $f_{st}(x)$ are taken into account. From (3) equations are written for the moments of any order of the random variable $q(t)$, describing the random variable of the stock $X(t)$ ($\frac{\partial^n F(e^{-\theta},t)}{\partial\theta^n}_{|\theta=0} = (-1)^n\int_0^\infty q^n\omega(q,t)dq = \boldsymbol{E}(q^n)$ are the moments of the random variable $X(t)$).

In Ref. [36] the equilibrium fluctuations of the internal thermodynamic parameters *B(t)* are specified by a stationary Markov process characterized by kinetic coefficients:

$$K_{\gamma_1..\gamma_n}(B) = \lim\nolimits_{\tau\to 0}[\tau^{-1}\left\langle \Delta B_{\gamma_1}...\Delta B_{\gamma_n}\right\rangle_B], \tag{20}$$

where $\Delta B_{\gamma_k} = B_{\gamma_k}(t+\tau) - B_{\gamma_k}(t)$, and averaging is performed at a fixed value of *B(t)*.

The first kinetic coefficient, characterizing the rate of change of the system state, in the storage model (1)–(2) is equal to (21):

$$K_1(B) = \rho - r_\chi(B); \qquad \rho =< \frac{\partial A}{\partial t} >= \frac{\partial\varphi(\theta)}{\partial\theta}_{|\theta=0}. \tag{21}$$

This value is also called the drift vector Refs. [1, 3, 37, 39]. It completely describes the evolution of the system in the deterministic case (macroscopic phenomenological equations of linear relaxation of the form (45)). If we proceed from the exact stochastic storage model, then for the Laplace transform we obtain equation (3).

Let us now consider the output model of the type:

$$r(x) = bx, \quad b > 0\cdot \tag{22}$$

At (22):

$$\partial F_{bq}(\exp\{-\theta\},t)/\partial t = [-\beta\varphi(\theta/\beta) - \theta b\partial/\partial\theta]F_{bq}(\exp\{-\theta\},t)\,. \tag{23}$$

The Laplace transform of the random variable *X(t)=B(t)* from (1), (2), and (22) has the form:

$$F_{bq}(e^{-\theta},t) = \boldsymbol{E}(\mathrm{e}^{-\theta B(t)}\,|B(0) = B_0) = \exp\{\text{-}\theta \mathrm{B}_0\mathrm{e}^{-bt} \text{ - }\lambda\int_0^t[1-\Psi(\theta e^{-bu})]\mathrm{du}\}\,;$$

$$\Psi(\theta) = \int_0^\infty \exp\{-\theta x\}\nu(dx)/\lambda = \int_0^\infty \exp\{-\theta x\}g(x)dx\,. \tag{24}$$

Random measure $\nu(dx) = \lambda g(x)dx = \lambda\mu\exp\{-\mu x\}dx$ for Eq. (17).

For the Laplace transform of the stationary probability density (18) for *a*=0 (7) of the random variable *B*, we obtain from (23)–(24) for $\partial \ln F/\partial t = 0$, that:

$$\ln F(\exp\{x\}) = -\int_0^{-x}(\varphi(u)/u)du/b; \quad A(x) = \varphi(-x)/b(-x)\,;$$

$$f(A) = \rho - bA(x) = dA/dt; \quad A(t) = \exp\{-bt\}(A_0 - A_{st}) + A_{st}; \quad A_{st} = \rho/b\,. \tag{25}$$

The stationary value of the quantity (2) is equal to (25) From (25) we find that $P_0 = 0,\, P_0(\exp\{-v\}) = \exp\{-\int_v^\infty \varphi(u)du/bu\} = 0$, i.e. the values of the probability density at the zero point are equal to zero, there is no jump and singularity of the distribution density at zero, as it should be at r(0+)=0.

The average lifetime $\left\langle\Gamma(x)\right\rangle$ of such a system is related to $P_0$ the expression [9]:

$$\Pi\{0\} = P_0 = [1 + \int_0^\infty \langle \Gamma(x) \rangle \nu(dx)]^{-1}, \quad (26)$$

and tends to infinity.

The stationary values $\langle B(\exp\{-v\}) \rangle$ obtained from (25) are equal to:

$$\langle B(\exp\{-v\}) \rangle = \varphi(v)/bv; \qquad \langle B(\exp\{-v\}) \rangle_{|v=0} = \rho/b; \quad \rho = \lambda\mu^{-1}. \quad (27)$$

## 2. Kinetic potential, its image, flows

For the kinetic coefficients (20) the function is written Ref. [36]:

$$\Phi(\theta, y) = \sum_{m=1}^{\infty} \frac{1}{m!} \sum_{\alpha_1 \ldots \alpha_m} K_{\alpha_1 \ldots \alpha_m}(y) \theta_{\alpha_1} \ldots \theta_{\alpha_m}. \quad (28)$$

In Ref. [36] the kinetic potential *V(y, B)* of the argument y is defined as the generating function for the coefficients $K_m(q) \sim \beta^{1-m}$ of the kinetic equation:

$$K_{\gamma_1 \ldots \gamma_m}(B) = (k_B T)^{m-1} [\partial^m V(y, B) / \partial y_{\gamma_1} \ldots \partial y_{\gamma_m}]_{|y=0}, \qquad m \ge 1; \quad (29)$$

the kinetic potential in Ref. [36] is written as:

$$V(\theta, B) = \sum_{m=1}^{\infty} \frac{1}{m!} \beta^{m-1} \sum_{\alpha_1 \ldots \alpha_m} K_{\alpha_1 \ldots \alpha_m}(B) \theta_{\alpha_1} \ldots \theta_{\alpha_m}. \quad (30)$$

The kinetic potential is related to the function Φ (28) by the relation:

$$\Phi_\beta(v, B) = \beta V(v/\beta, B), \quad (31)$$

where $\beta^{-1}$ is a small parameter; for an equilibrium Gibbs system $\beta = 1/k_B T_{eq}$, $k_B$ is the Boltzmann constant, $T_{eq}$ is the equilibrium temperature.

The introduction of the parameter $\beta$ creates the prerequisites for the physical interpretation of various stochastic systems. Thus, in the equilibrium canonical Gibbs ensemble $\beta = 1/k_B T$, where $k_B = 1.38 \times 10^{-23}$ J/K is the Boltzmann constant, T is the absolute temperature. The Boltzmann constant is a small value, this is used in Ref. [36]. In the general case, the value $\beta$ takes into account the influence of the environment. We consider it a small parameter $\beta^{-1}$. In a number of works (for example, Ref. [36]), stationary nonequilibrium states are considered, where the role of the value $\beta^{-1}$ in (31) is played by the noise level $\kappa \ge 0$ or the intensity of the external source of fluctuations. The value $\beta^{-1}$ and $\kappa \ge 0$ characterize the ratio of microscopic and macroscopic scales. The limiting case $\beta^{-1} \to 0$ corresponds to deterministic behavior.

The kinetic potential *V(-y, q(t))*, determined by formula (30) for equation (1), is equal to:

$$V(-y, q(t)) = -\varphi(y) + y r_\chi(q(t)); \quad r_\chi(q(t)) = r(q(t)) - r(0+)\chi_{q(t)}; \quad \boldsymbol{E}\{q(t)\} = P\{q(t) = 0\} = P_0(t),$$

$$\varphi(\theta) = \int_0^\infty (1 - exp\{-\theta x\}) \nu(dx); \quad \nu(dx) = \lambda g(x) dx; \quad \varphi(\theta) = \lambda - \lambda\psi(\theta);$$

$$\Phi_\beta(y, B) = -\varphi_\beta(-y) - y r_\chi(B); \quad \lambda_\beta = \lambda\beta, \; g_\beta(x) = \beta g(\beta x), \; \varphi_\beta(y) = \beta\varphi(y/\beta), \quad (32)$$

where r(q) is the continuous part of the output function, $P_0(t) = P_{q_0}(0, t) = E\{\chi_{q(t)}\} = P\{q(t) = 0 | q(0) = q_0\}$ is the probability of system degeneration.

The direct kinetic equation for the distribution density of the number of elements in the system p(y, t) has the form Ref. [36]:

$$\frac{\partial p(y,t)}{\partial t} = N_{\partial, y} V(-\frac{1}{\beta} \frac{\partial}{\partial y}, y) p(y,t), \quad (33)$$

where the operator $N_{\partial,y}$ determines the order of operations (the operation $\partial / \partial y$ is performed after multiplication by y). The Laplace transform of the relation (33) gives an equation for the Laplace transform of F of the form $E(e^{-\theta X}) \equiv F(e^{-\theta}) = \int e^{-\theta y} p(y,t)dy$ of the form (3) through (32),

$$\frac{\partial F(e^{-\theta},t)}{\partial t} = N_{\theta,\partial/\partial\theta}\Phi(-\theta,-\frac{\partial}{\partial\theta},t)F(e^{-\theta},t), \tag{34}$$

coinciding with (3) for V(-y, q(t)) of the form (32).

For the stochastic potential $\Phi_\beta(y,q)$ (31)-(32) of the storage model (1), the expression for the first kinetic coefficient is (29), (21):

$$K_1(q) = \partial\Phi_\beta(y,q)/\partial y_{|y=0} = \rho - r_\chi(q). \tag{35}$$

This expression does not depend on β. Averaging expression (35) over distribution (43) leads to the expression:

$$\kappa_1(x) = \int K_1(q)\omega_x(q)dq = \rho - \varphi(-x)/(-x). \tag{36}$$

The average kinetic coefficients $\kappa_n = \int K_n(q)\,\omega_x(q)dq,\ n \geq 2$, do not depend on q in the storage model and are not affected by averaging over (43):

$$\kappa_2 = \beta^{-1}\partial^2\varphi(y)/\partial y^2{}_{|y=0};\quad \kappa_3 = \beta^{-2}\partial^3\varphi(y)/\partial y^3{}_{|y=0};\ \ldots\quad \kappa_n = \beta^{1-n}\partial^n\varphi(y)/\partial y^n{}_{|y=0}. \tag{37}$$

In Refs. [22-23] a generalization of this approach was carried out.

### 3. Conditions of stationary of the storage model

In the work [1], it is noted that in a statistical system a phase transition (including a nonequilibrium one) occurs if the stationary states of the system, describing random variables, undergo qualitative changes. For storage processes (1), several criteria can be specified that stationary states exist. Some of them are associated with lifetimes (4) and with the times of reaching lower levels. Others are associated with input and output parameters. Thus, in the work [9] a necessary and sufficient condition for the existence of stationary states of process (1) is obtained. These states are possible if there is a value of $w_0$ for which the inequality is satisfied:

$$\sup_{w \geq w_0} \int_0^\infty \int_w^{w+y} (r(u))^{-1} du\, \nu(dy) < 1, \tag{38}$$

where *v* and *r(u)* are expressions from relations (1)-(2), (32), w is some sufficiently large value, *ν(y)* is the Levy measure Refs. [2, 5], it is included in the cumulant $\varphi(s)$ of the process *X(t)* (2), (32); for Eq. (17) $\nu(dx) = \lambda g(x)dx = \lambda\mu\exp\{-\mu x\}dx$. If condition (38) is violated, there are no stationary states, there are no distributions of the form (18), relations (27), (44)-(46) written for non-stationary states derived from stationary states by the action of some thermodynamic forces and relaxing to stationary states are not valid. If condition (38) is violated and stationary distributions are absent, there are no stationary structures corresponding to the macroscopic quantities characterizing the physical system. In this case, the lifetime (4) is not defined, and the lifetime in such a system is determined by the stationary environment. Such a system can collapse, and new structures for which condition (38) is valid can be formed in it. Physically, this corresponds to a phase transition with a change in the structure of the system. This approach to describing phase transitions differs from the approach used, for example, in [1], which is explained by the use of different basic stochastic models. Under $\boldsymbol{E}A(t) < \infty$ condition (38) is equivalent to $\sup_{w \geq w_0} \boldsymbol{E}A(t)/r(w) < 1$.

The connection between the finiteness of the mean lifetime $\langle \Gamma(x) \rangle$ and the existence of stationary distributions and non-zero values of the stationary probability of degeneration $P_0 = \lim_{t\to\infty} P_0(t)$ is evident from equality (26), which follows from the relation:

$$P_0(s) = \int_0^\infty P_0(t) e^{-st} dt = \frac{\boldsymbol{E}(x,s)}{s + \lambda - \lambda \int dy g(y) \boldsymbol{E}(y,s)}; \quad \boldsymbol{E}(x,s) = \int_0^\infty e^{-st} p(\Gamma_x \le t) dt . \tag{39}$$

Here ***E**(x,s)* is the Laplace transform of the lifetime, $P_0=1/Q$, $Q$ is the equilibrium partition function, $\lambda$ is the rate of receipts (2), *g(x)* is the distribution from Eq. (2).

The properties of the stationary distributions $\Pi$ of process (1) depend on the properties of the function $r$ from (1) and the measure $\nu$ from (2), (24), (32). For continuous values of $r$ and $\nu$, the values $\Pi(dy)$ are continuous and have an atom at zero, that is, as in (9), (18), $\Pi(dy) = [P_0\delta(y) + f(y)]dy$ , where *f(y)* is a continuous function.

For $P_0>0$, when $r$(0+)>0, the equation for the function *f(x)* is written as [6-9]:

$$r(x) f(x) = P_0 \nu(x,\infty) + \int_0^x \nu(x-y,\infty) f(y) dy; \quad \int_0^\infty e^{-\alpha x} \nu(x,\infty) dx = \varphi(\alpha) / \alpha ;$$

$$\nu(x,\infty) = \lambda(1 - \int_0^x g(y) dy); \qquad \int af(x) e^{-\nu x} dx = a(F(e^{-\nu}) - P_0) . \tag{40}$$

For model (9), the necessary and sufficient conditions for the existence of stationary distributions (38) are written as:

$$sup_{w \ge wo} \int_0^\infty log\left[1 + by / \left(a + bw\right)\right] \nu\left(dy\right) / b < 1. \tag{41}$$

For $by << a$, we obtain from (41) using a series expansion a coincidence with the condition for the existence of stationary distributions for the model $r=a$. For $by >> a$, we obtain from (41) a condition that coincides with (42). The relationships for the flows are written down. It is possible to carry out a physical interpretation of the quantities $a$, $b$, $\varphi(s)$.

The condition for the existence of stationary distributions for the model (22), obtained from (38), has the form:

$$\int_1^\infty lnx \nu\left(dx\right) < \infty, \tag{42}$$

and is fulfilled for almost all real systems, i.e. such a system always has stationary states.

## 4. Fictitious thermodynamic forces and the evolution of the system

We use the approach proposed in Ref. [36] to describe external influences on the system, applying it to stochastic storage processes. This approach considers open systems in which nonequilibrium stationary states are possible, Ref. [36]. Such states, rather than equilibrium ones, are chosen as the basic states. An auxiliary distribution of the form is introduced:

$$\omega_x(B) = e^{\beta x B} \omega_{st}(B) / F(e^{\beta x}), \quad F(e^{\beta x}) = \int_0^\infty e^{\beta x B} \omega_{st}(B) dB , \tag{43}$$

corresponding to the action of virtual thermodynamic forces x, deviating the state of the system from a stationary nonequilibrium state with distribution $\omega_{st}$ , Ref. [36]. These forces can also be real. Real thermodynamic forces $h$ can be introduced by replacing the value of $x$ in (43) with $x - h$, Ref. [36]. Distribution (43) was obtained in Ref. [36] from the extremum of the Kullback entropy for fixed average values of the parameters $B$, $\langle B \rangle = A$ , and the Lagrange multiplier equal to $-x$. If the forces $x$ in (43) are virtual, then distribution (43) is nonequilibrium. It is used to describe relaxation processes in the system.

If we denote the mean value $\langle B \rangle = A$ and define the dependence *x(A)* as an inverse dependence in the ratio:

$$\frac{\partial \ln F(e^{\beta x})}{\partial x} = \frac{\beta \int B e^{\beta x B} \omega_{st}(B) dB}{\int e^{\beta x B} \omega_{st}(B) dB} = \beta A(x) , \tag{44}$$

where *A(x)* denotes the average value of the quantity *B* under the acting forces *x*, the Laplace transform $F(e^{\beta x})$ is taken from the expression (43), then substituting the expression *x(A)* found in this way into relation (36) allows us to compare the obtained result with the phenomenological equation for the average value *A* of the form:

$$\frac{dA}{dt} = f(A) . \tag{45}$$

Averaging expression (35) over distribution (43) in the stationary case leads to expressions (36), (46),

$$\frac{d\langle B \rangle}{dt} = \kappa_1(x) = \int K_1(B) \omega_x(B) dB ; \qquad \kappa_1(x) = \rho - \varphi(-x)/(-x) , \tag{46}$$

where it is used that in the stationary case $\langle r_{-v} \rangle = \varphi(v)/v;\ v = x/\kappa$. Changes in the system under consideration are described by the phenomenological (macroscopic) equation (45), where the right-hand side of (45) is written in accordance with expressions (46), (44). This means that the derivative $\dot{A}$ in (45) is understood as the result of averaging the derivative $\dot{B}$ (*B* is a random variable), as in (46), and the averaging is carried out not at a fixed value of *A*, but at fixed forces *x(A)*.

For a single-phase system with a large number of particles and small fluctuations, the distribution (43) is sharp and concentrated near the point $A(x) = \int B \omega_x(B) dB$, and $x = \partial \Psi(B)/\partial B$, where $\Psi(B)$ is an analogue of the conditional free energy of the equilibrium Gibbs distribution for the stationary nonequilibrium case Ref. [36]. Expressions (43), (35), (46), (44)–(45) establish a correspondence between the nonequilibrium average:

$$A(t) = \int B \omega(B,t) dB$$

by the non-stationary distribution *ω(B,t)*, depending on time, and the average by the nonequilibrium, but stationary distribution (43):

$$A(x) = \int B \omega_x(B) dB ,$$

i.e. a correspondence is established:

$$\omega(B,t) \sim \omega_{x(A)}(B)$$

between the time-varying distribution function *ω(B, t)* and the distribution (43), the changes in which are assumed to be stationary, occurring due to changes in the forces x(A). Following Ref. [36], we approximate the non-stationary behavior of the system by a quasi-stationary distribution. In this case, the value of (43) is expressed in the current approach as known and initial. Thus, time changes can contribute to the introduction of virtual stationary thermodynamic forces *x*.

The behavior of the stochastic storage system (1) is related to the type of the output function $r_\chi[X(t)] = r(X(T)) - r(0+)\chi_{X(t)}$. For the exit velocity at zero r(0+)=0, the probability of degeneration $P_0=P_0(exp\{-v\})=0$ for any impacts $v=x\beta$. The average lifetime of the system in this case tends to infinity. The system's equilibrium is not achieved, but stationary nonequilibrium states are possible. In this case, a general pattern is observed: the finiteness of the system's lifetime corresponds to its nontrivial macroscopic behavior, depending on the characteristics of the system's interaction with the environment included in the value of $\beta$ (and in other parameters of the model). And vice versa: in systems with an infinite lifetime, in which *r(0+)=0*, the average values behave in a standard way. In general, establishing a one-to-one correspondence between a mathematical model and the area of physical problems it describes is a special, extensive subject of research.

For the first kinetic coefficient with thermodynamic forces for the model *r=a=1* (7) we obtain:

$$K_1(N_1,q)=I(q,\ t_1)=lim_{\tau\to 0}E[(N_{t_1+\tau}-N_{t_1})\ exp\{-q(N_{t_1+\tau}-N_{t_1})\}\,|\,N_{t_1}]/\tau=\rho_\theta-1+\chi_{N_1};$$

$$\rho_\theta=\partial\varphi(\theta)/\partial\theta;\qquad \partial V(-y,N)/\partial y=-K_1(N,y\beta);\qquad \theta=y\beta. \tag{47}$$

Averaging (47) over distribution (43), we obtain the flow values:

$$<I_{No}(exp\{-v\},\theta,t_1)>=\int_0^\infty K_1(N_1,q)\omega_{-v}(N_1\,|\,N_0)dN_1=\rho_\theta-1+P_{No}(0,t_1)/F_{No}(exp\{-v\},t_1); \tag{48}$$

$$<I(exp\{-v\},\theta,t)>=\int_0^\infty K_1(N,\theta)\omega_{-v}(N)dN=\rho_\theta-1+P(0,t)/F(exp\{-v\},t). \tag{49}$$

Averaging (32) with the output of type (7) according to (43) leads to expressions for the representation of the kinetic potential, Ref. [36]:

$$R_{No}(-y,-v)=\int_0^\infty\Phi_t(-y,N)\omega_{-v}(N\,|\,N_0)dN=-\varphi(y)+y\left[1-P_{No}(0,t)/F_{No}(exp\{-v\},t)\right]; \tag{50}$$

$$R(-y,-v)=\int_0^\infty\Phi_t(-y,N)\omega_{-v}(N)dN=-\varphi(y)+y\left[1-P(0,t)/F(exp\{-v\},t)\right]. \tag{51}$$

Using (47) in (32) and (48)-(49) in (50)-(51) gives:

$$\Phi_t(-y,N)=-\varphi(y)+y[\rho_\theta-K_1(N,\theta)]; \tag{52}$$

$$R_{No}(-y,-v)=-\varphi(y)+y[\rho_\theta-<I_{No}(exp\{-v\},\theta,t)>]; \tag{53}$$

$$R(-y,-v)=-\varphi(y)+y[\rho_\theta-<I\{exp\{-v\},\theta,t)>]. \tag{54}$$

Averaging $\Phi_t$ (32) and $K_1$ (35) for arbitrary output functions over the distribution (43) leads to expressions for the representation of the kinetic potential Ref. [36]:

$$R_t(-y,-v)=\int_0^\infty\Phi_t(-y,N)\omega_{-v}(N)dN=-\varphi(y)+y<r^0{}_{-v}(t)>; \tag{55}$$

$$<r^0{}_{-v}(t)>=\int_0^\infty r_\chi(N(t))\omega_{-v}(N,t)dN=<r(exp\{-v\},t)>-r(0+)P(0,t)/F(exp\{-v\},t);$$

$$<r(exp\{-v\},t)>=\int_0^\infty r(N)exp\{-vN\}\omega(N,t)dN/F(exp\{-v\},t);$$

and flows:

$$<I(exp\{-v\},t)>=\int_0^\infty K_1(N)\omega_{-v}(N,t)dN=\rho-<r^0{}_{-v}(t)>=-\partial R(-\theta,\ -v)/\partial\theta_{|\theta=0}. \tag{56}$$

From (56) we obtain that:

$$<I(exp\{-v\},t)>=\rho-[R_t(-y,-v)+\varphi(y)]/y. \tag{57}$$

In the stationary case, the dependence of flows on impacts has the form:

$$\langle I_{stv}\rangle=\rho-\varphi(v)/v,$$

and:

$$R(-y,-v)_{|y=v}=R(-v,-v)=0;\qquad v=X/\kappa, \tag{58}$$

where $X$ is the stationary thermodynamic forces, $\kappa\geq 0$ is the parameter characterizing the noise intensity [36], equal to $k_BT$ if the noise intensity is approximately the same as in the equilibrium state.

That is, in the stationary case:

$$< r^0_{-v} >= \varphi(v) / v; \qquad v = X / \kappa; \qquad R_{st}(-y,-v) = -\varphi(y) + y\varphi(v) / v. \qquad (59)$$

In the works [36] the value *R* (50) - (51), (53) - (55) corresponds to the generating function for the diffusion coefficients:

$$D_n(x,X) = \int K_n(B,X) exp\{-\beta xB\}\omega(B) dB / F(exp\{-\beta x\}),$$

$$D(u, x) = \sum_{n=1}^{\infty} u^n D_n(x) / n! = \tau^{-1} lnSpexp\{ln\rho(x) + u_\alpha \int_0^\tau \hat{I}_\gamma(t) dt\},$$

$$\hat{I}_\gamma = \dot{B}_\gamma = Y_\gamma(B) \equiv Sp\{\rho(x) i[\hat{H}_0, \hat{B}_\gamma] / \hbar\}, \qquad (60)$$

where $\rho(x)$ is the nonequilibrium distribution, $\hat{H}_0$ is the unperturbed Hamiltonian, *x* and *X* are thermal and dynamic perturbations, Refs. [41, 42].

From (56)-(59) we find that for the relation (22),

$$< r^0_{-v} >= b < I(exp\{-v\}) >; \quad < I(exp\{-v\}) >= b(< q > - < q(exp\{-v\}) >); \quad < q >=< q(exp\{-v\}) >_{|v=0}. \quad (61)$$

As macroscopic variables <*q*> we can choose flows <*I*>, as in the case of stochastic description of stationary nonequilibrium open systems;

The general stochastic equation for the contents of some storage facility in inventory theory is (1). This relationship is a Langevin-type equation. Using the value *r(x)* various situations can be described. Thus, when

$$r(x) =< dA / dt > +\alpha x + \beta x^3 - \gamma\Delta x, \qquad (62)$$

substitution of (62) into (1) leads to the Landau-Ginzburg equation with a random force $F=-<dA/dt>+dA/dt$. The term $\gamma\Delta x$ can be considered as an average value rather than a random variable. For $r(x)=<dA/dt>+\alpha x$ we obtain the usual Langevin equation.

The physical meaning of stochastic storage theory has been used many times, for example, in general systems theory. It consists of elements entering and leaving the system.

If there are several components $q_\alpha$, $\alpha=1,\dots, n$, then we can write a system of equations:

$$dq_\alpha / dt = dA_\alpha / dt - r[q_1,\dots,q_n], \qquad \alpha = 1,\dots, n,$$

i.e. the quantities $q_\alpha$ are related via the output functions. For example, if $r(x)=bx$, then for the multidimensional vector $\vec{q}=(q_1,\dots,q_n)$ we can introduce a square matrix $b_{ij}$, the quantities $\vec{\varphi}(\ )=(\varphi_1(),\dots, \varphi_n())$ and $\hat{\varphi}^T(\vec{\theta})=\varphi_1(\theta_1)+\varphi_2(\theta_2)+\dots+\varphi_n(\theta_n)$. The kinetic potential is equal to:

$$V(-\vec{\theta},\vec{q}) = \vec{\theta}^T \hat{b}\vec{q} - \hat{\varphi}^T(\vec{\theta})$$

(in component-wise notation $\theta_i b_{ij} q_j - \varphi_i(\theta_i)$). Equation (3) in the stationary case takes the form $0 = -\theta_i b_{ij} \partial F(exp\{-\theta\}, t)/\partial\theta_j - \varphi_i(\theta_i) F(exp\{-\theta\}, t)$.

If we look for *F* in the form $exp\{A(\theta)\}$, then $\theta_i b_{ij} \partial A / \partial \theta_j = -\varphi_i(\theta_i)$, and:

$$\partial A/\partial\theta_j = \sum_{k=1}^{n} \tilde{b}_{jk} \hat{\varphi}^T(\vec{\theta}) / n\theta_k,$$

where $\tilde{b}$ is the inverse matrix, $\hat{\varphi}^T(\vec{\theta})$ is a scalar.

Expression (32) has a macroscopic meaning. If, using the relation (31) $\Phi_\beta(\theta,y)=\beta V(\theta/\beta,y)$, which connects the kinetic potential *V* with the stochastic potential $\Phi_\beta$ (it is not of a macroscopic nature) and defines a family of stochastic potentials $\Phi_\beta$, depending on $\beta$, we introduce a family of stochastic potentials:

$$\Phi_\beta(-y, q(t)) = -\varphi_\beta(y) + yr_\chi(q(t)), \qquad (63)$$

then only the component $\varphi_\beta(y)$ changes; in this case, relations (32) are satisfied for $\varphi_\beta(y)$, $\lambda_\beta$, $b_\beta$.

The value $\rho=<\partial A/\partial t>=\partial\varphi(\theta)/\partial\theta_{/\theta=0}$ does not change. The criterion for the termination of the expansion of the value $\varphi(\theta)$ over $\theta$,

$$\varphi_\beta(-\theta)=\lambda\beta\int_0^\infty(1-exp\{\theta x\})\beta b(\beta x)dx=\lambda\beta\int_0^\infty(1-exp\{\theta y/\beta\})b(y)dy=$$
$$-\theta\rho-\theta^2\sigma^2/2!\beta-\theta^3\lambda<x^3>/3!\beta^2+...;$$
$$\sigma^2=\int_0^\infty y^2b(y)dy;\qquad \rho=\lambda<x>;\qquad <x>=\int_0^\infty xb(x)dx;\qquad <x^3>=\int_0^\infty x^3b(x)dx,$$
there will be small values of $\beta^{-1}$, for example, high temperatures. A decrease in temperature gives an increase in the contribution of jumps and the need to take into account higher-order moments. Expressions (32), (63) are valid for any $\beta$.

**5. Macroscopic equations for the storage model in the approximation of steady-state thermodynamic forces.**

Let us apply to the storage model (1) relations (43), (44)-(46), establishing a correspondence between the nonequilibrium average macroscopic state of the stochastic model and the average value obtained by using averaging over a distribution containing stationary virtual thermodynamic forces. In this way, general relations (35)-(37) are obtained. By specifying explicit expressions for the density of the distribution function of the magnitude of jumps in receipts, we can find the function φ(s) included in (36)-(37). Let us consider a distribution of the form (17) and a constant output (7).

We can assume $a=1$, replacing $\lambda$ with $\lambda/a$ and scaling time. The first kinetic coefficient (21) in this case is equal to $K_1(B)=\rho-1+\chi_B$. Averaging this expression with the distribution (43) leads to the expression (see also (47)):

$$\kappa_1(x)=\rho-1+P_0/F(\exp\{\beta x\});\quad F\{\exp\{\beta x\}\}=P_0[1-\varphi(-x)/(-x)]^{-1},$$
$$P_0=1-\rho/a;\quad P_0=0,\quad \rho\geq a;\quad P_0=lim_{t\to\infty}P_0(t), \tag{64}$$

where (46) is used. For a function $g(x)$ of the form (17) $\rho=\lambda/\mu;\quad \varphi(\theta)=\lambda\theta/(\mu+\theta)$.

From relations (44), (64) we find that when $\langle B\rangle=A$,

$$\beta A(x)=\lambda/(\mu-x)(\mu-\lambda-x). \tag{65}$$

The inverse dependence $x(A)$ is found from the quadratic equation. Depending on the value of $x$, the behavior of $\beta A(x)$ can be divided into three regions. In the region of $x$ values $\mu-\lambda<x<\mu$, there are no physical values of $\beta A(x)>0$, which can be interpreted either as the inapplicability of the distribution (43) in this region as an equivalent of the true distribution $\omega(B, t)$, or as the destruction of this hierarchical level $B$ at these values of $x$. Real thermodynamic forces $h$ can enter into expression (43) in the same way as fictitious forces $x$, more precisely, in combinations $x$-$h$, Ref. [36]. Therefore, with the latter interpretation, it can be considered that some action within certain limits destroys the system. A constant exit velocity leads to a singularity at zero, a nonzero probability of degeneracy, and the presence of several branches of the system's evolution.

From (65) we find that:

$$x=\mu-\lambda/2a\mp\lambda(1+4a/\lambda\beta A)^{1/2}/2a, \tag{66}$$

where the signs + and – correspond to the branches of $\beta A(x)$. For one of the branches:

$$x=\mu-\lambda/2a-\lambda(1+4a/\lambda\beta A)^{1/2}/2a. \tag{67}$$

As an example, one can point to the behavior of the neutron number density in nuclear reactors under certain influences that change the effective multiplication factor $k_{ef}$; at $k_{ef}\to 1$, the average neutron number density tends to infinity. In regions, where $x>\mu$, the integral for $\varphi(x)$ diverges and the values of branch $A^-$ have no physical meaning.

Substituting (67) into (45), we obtain, using (46), that:

$$f(A)=\rho-2a/[1+(1+4a/\lambda\beta A)^{1/2}]\,. \qquad (68)$$

The steady-state value of $A$ is determined from (44) at $x=0,\, A_{st}=A(x=0)$, and is equal to $\beta A_{st}=\sigma^2/2(1-\rho);\, \sigma^2=2\lambda/\mu^2$. Substituting $A_{st}$ into (68) yields $f(A_{st})=0$. At $\beta\to\infty$ we obtain a low-temperature (or low-noise) approximation, which for model (7) has the form:

$$dA/dt=\rho-1\,. \qquad (69)$$

The solution to this equation (69): $A_{\beta\infty}=A_0-(1-\rho)t$ The value of $A_{\beta\infty}$ at $\rho<1$ reaches zero in time $t_{\max}=A_0/(1-\rho)$, which coincides with the average value of the lifetime $<\Gamma_{Ao}>$ for the storage model $r=1$. At $\rho<1$, the behavior $A_{\beta\infty}$ is stable. At $\rho>1$, it is unstable. An analogy can be drawn with stable and unstable modes. Relation (69) is valid not only for *g(x)* of the form (17), but also for all *g(x)* and *φ(s)*, since:

$$f(A)=\rho-1+aP_0\exp\{-\beta\int_0^x A(y)dy\}_{|\beta\to\infty}\to\rho-1\,.$$

It is possible to introduce a potential and describe phase transitions. It is also possible to carry out expansions in series of the function *f(A)* in powers of $\beta^{-1}$ and around the values of $A_{st}$. The stationary condition (38) for the model $r=a$ takes the form: $\rho/a<1$. In region, where $\mu-\lambda<x<\mu$, replacing the distribution $\omega(q,t)$ with the distribution $\omega_x(q)$ is invalid, since this is possible for relaxation processes, and there are no stationary states in this region; in Ref. [11] it is shown that for the model $r$=1, the effect on the system can be taken into account by replacing the value of $\rho$ from (64) with $\rho_x=\varphi(-x)/(-x)$; in this case it turns out that for $x>\mu-\lambda$, $\rho_x>1$, i.e. the condition for the existence of stationary states is violated.

Expanding the function $f^+(A)$ from (68) into a series in powers of $\beta^{-1}$, we obtain:

$$f(A)=\rho-1+1/\lambda\beta A-2/(\lambda\beta A)^2+3/(\lambda\beta A)^3+\ldots\,.$$

It is possible to expand around the values of $A_{st}$. Since of $f(A_{st})=0$, then:

$$f(A)=\frac{\partial f(A)}{\partial A}_{|A_{st}}(A-A_{st})+\ldots=\frac{\rho^3}{(2-\rho)(\rho-1)^2\lambda\beta A_{st}{}^2}[A-A_{st}-(1-\frac{1}{\lambda\beta A_{st}}\frac{\rho(2\rho-3)}{(2-\rho)(1-\rho)})(A-A_{st})^2]+\ldots.$$

If the output function in the storage model is equal to $r(x)=bx$ (22), then for the Laplace transform of the stationary probability density of the random variable $q$ we obtain (25).

Expressions (25) are valid for any values of $\beta$ and *φ(s)*. In this respect, they coincide with the low-noise approximation written for model (7) in (69) under the assumption that $\beta^{-1}\to 0$. In this case, there is no dependence on $\chi_q$, $r(0+)=0$, the degeneration probability $P_0=P_0(exp\{-v\})=0$ for any effects $v=x\beta$. The average lifetime of the system in this case tends to infinity. Thus, a general pattern is observed: the finiteness of the lifetime of the system corresponds to its nontrivial macroscopic behavior, depending on the characteristics of the interaction of the system with the environment included in the value of $\beta$. And vice versa: in systems with an infinite lifetime, in which $r(0+)=0$, the average values behave in a standard way. For (17) and (22):

$$F(exp\{-\theta\})=(1+\theta/\mu)^{-\lambda/b}\,.$$

This Laplace transform corresponds to the stationary probability density (18) (for $a$=0). The behavior of $P_{st}$ (or $\omega_{st}$) for $\lambda<b$ and $\lambda>b$ is shown in Fig. 1. For $\lambda=b$, the external noise-induced phase transition Ref. [4] occurs. Since, in accordance with (32), $\lambda_\beta=\lambda\beta$, this transition in the storage model (1), (22) corresponds to the transition for $\beta=\beta_c$. If we add a small constant value $a<<\rho$ to the output function (22) (in this case we obtain case (9)), then the system will have a finite lifetime (this follows from (26)), and it will be able to reach the zero point. For $a\to 0$, the average lifetime $<\Gamma>\to\infty$ according to the scaling law. It is possible to determine the critical indices in the storage model by considering, for example, the temperature to be a quantity proportional to $\beta^{-1}$, and the order parameter to be the quantity a from (9). Then the critical indices are determined at $a$=0. The phase transition induced by external noise occurs at $a=\rho$. In the general case, we can speak of a group of transformations of the storage model, a particular example of which is the transformation with respect to $\beta$ (22). Similar transformations can be carried out for the external field associated with the argument $\theta$ (or with $x$ from $\theta=-\beta x$) and for the correlation time $\tau$ - a measure of the distance from white noise (in the model (9) $\tau_{corr}\sim 1/b$, and it is necessary to determine the families $b_\varepsilon=\tau_b$). The storage model is characterized by a point in the parameter space $(\beta,\ \theta,\ \tau)$. For the model (9), explicit expressions for the behavior of the average values are also written, combining the cases considered.

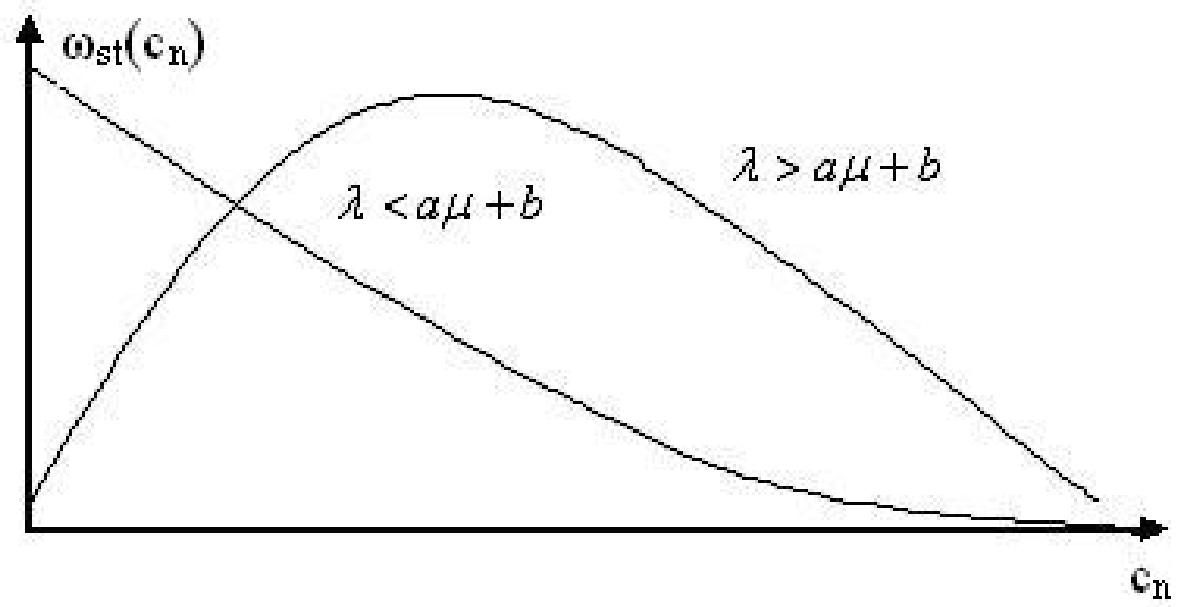

**Fig 1**. Phase transition: at $\lambda = a\mu + b$ the distribution maximum emerges for different ratios between the input intensity $\lambda$ and the output velocity $b$.

### 6. Macroscopic relations for exact stochastic storage equations

For the stochastic potential (32) of the storage model (1), the expression for the first kinetic coefficient is equal to (35).

This expression does not depend on β. Averaging expression (35) over distribution (43) leads to expression (36), where the ratio (32) is used $(\lambda_\beta=\lambda\beta;\ g_\beta(x)=\beta g(\beta x);\ \varphi_\beta(y)=\beta\varphi(y/\beta))$.

For values of $x$ and $\beta$ independent of $t$, the solution of equation (8) yields that:

$$F_a(exp\{-\theta\},t) = exp\{[-\beta\varphi(\theta/\beta)+\theta a]t-\theta q_0\} -$$

$$a\theta\int_0^t P_0(\tau)exp\{(t-\tau)[-\beta\varphi(\theta/\beta)+\theta a]\}d\tau; \qquad F_a(exp\{-\theta\},t=0)=exp\{-\theta q_0\};$$

$$F_a(exp\{\beta x\},t) = exp\{-[\varphi(-x)+ax]\beta t+\beta x q_0\} - a\beta x\int_0^t P_0(\tau)exp\{-\beta(t-\tau)[\varphi(-x)+ax]\}d\tau;$$

$$F_a(exp\{\beta x\},t=0)=exp\{\beta x q_0\}; \quad q_0=q(t=0). \qquad (70)$$

In the stationary case, when $\partial \ln F_a(exp\{\beta x\},t)/\partial t=0$, from (8) we obtain expression (64).

The solution of the equation for $<q(x,t>=\partial \ln F_a(exp\{\beta x\},t)/\partial(\beta x)$, obtained from (8), gives:

$$<q(x,t> = exp\{-\int_o^t [ba\, xP_0(t)/F_a(exp\{bx\},\tau)]d\tau\}\{A_{t=0}(x)+\int_0^t [aP_0(t)/\ F_a(exp\{bx\},\tau)]d\tau +$$

$$\int_9^t (-a+\partial\varphi(-x)/\partial(-x))exp\{bax\int_0^\tau [P_0(u)/F_a(exp\{bx\},\ u)]du\}d\tau]\}.$$

The Laplace transform of expression (70) with respect to time is equal to:

$$F_a(exp\{\beta x\},s)=\int_0^\infty exp\{-st\}F(exp\{\beta x\},t)dt =$$

$$[exp\{\beta xq_0\} + \beta xaexp\{-q_0\eta_\beta(s)/a\eta_\beta(s)\}]/[s+\beta(ax+\varphi(-x))]; \quad (71)$$

$$a\eta_\beta(s) = s + \beta\varphi(\eta_\beta(s)/\beta); \qquad P_0(s) = \int_0^\infty exp\{-st\} P_0(t)dt = exp\{-q_0\eta_\beta(s)/a\eta_\beta(s).$$

For the stationary case, the expression for $F_{ast}(\exp\{\beta x\}) = \lim_{s\to 0} sF_a(\exp\{\beta x\}, s)$ coincides with (64). For x=0, the equation for <q(t)> has the form:

$$d<q>/dt = aP_0(t) - aP_0; \qquad <q(t)> = q_0 + a\int_0^t P_0(\tau)d\tau - aP_0 t. \quad (72)$$

For $< q_{st} >$ and $<q^2_{st}>$ from (71) we obtain:

$$<q_{st}> = lim_{s\to 0} <q(s)> = \sigma^2/2aP_0\beta; \qquad \rho < a; \qquad <q_{st}> \to \infty, \qquad \rho \geq a;$$

$$<q^2_{st}> = \lambda <x^3> /3aP_0\beta^2; \qquad \rho < a. \quad (73)$$

For the output model of the form $r(x)=a(1-\chi_x)+bx$ (9) from equation (3) we obtain Eq. (10). In the stationary case, equation (10) has a solution (11). The solution of the equation for $<q(t)>$ obtained from (10) is of the form (12).

For the output function $r(q)=a(1-\chi_q)+bq+cq^2$, the term $-cx\partial^2 lnF(exp\{\beta x\},t)/\beta\partial x^2$ is added to the right-hand side of equation (10), and the stationary equation takes the form

$$c\partial^2 F_{st}(exp\{\beta x\})/\partial x^2\beta + b\partial F_{st}(exp\{\beta x\})/\partial x + \beta(a-\varphi(-x)/(-x))F_{st}(exp\{\beta x\}) - a\beta P_{0\beta} = 0. \quad (74)$$

The equation for $<q>=A$ is written as:

$$\partial <q>/\partial t = -(a-\rho) + aP_0(t) - b<q> - c<q^2>,$$

i.e. the value of $<q>$ is expressed through $<q^2>$, we obtain a linked chain of equations, since $<q^2>$ will be expressed through $<q^3>$, etc. This chain can be uncoupled by solving equation (74), determining the value of $F_{st}(exp\{\beta x\})$, and then obtaining an equation for $x$ from $A=\partial F_{st}(exp\{\beta x\})/\partial(x)/\beta$. Substituting the obtained value $x(A)$ into (46) yields expression (45), i.e. uncoupling of the infinite chain (74) using relation (46). The same procedure should be followed for $r(q)=a(1-\chi_q)+bq+cq^2+dq^3+\ldots$. Expression (46) is written after averaging (35) using relations of the type (74), where $\partial^n F_{st}(exp\{\beta x\})/\partial(\beta x)^n/F_{st}=<q^n(x)>$. We do not consider cases of outputs of the type (74), since the main qualitative differences between noise of the type $dA/dt - <dA/dt>$ and Gaussian noise are obtained for lower powers of $r(q)$.

## 7. Phase transitions in the stochastic storage model

Below we consider phase transitions that arise in the stochastic storage model, similar to noise-induced transitions Ref. [1], when the macroscopic behavior of the system changes qualitatively under the influence of random external influences.

The properties of stochastic distributions are most easily expressed through the transition probabilities of the Markov process or its kinetic potential (30), (31), with the help of which the kinetic equation is written in the form (33), (3). The kinetic potential of the storage model is equal to (32). Random variables are described by the stochastic equation (3), and in the storage model they characterize the stock value. Let us specify the explicit form $g(x)=\mu\, exp\{-\mu x\}$ (17) for the distribution of the magnitude of input flow jumps. The continuous part $f(x)$ of the stationary distribution $P_{st}(x)=P(x=0)+f(x)$ of the stock value $x$ is determined from equation (40). The measure $\nu(x,\infty) = \int_x^\infty \nu(dx)$ for g(x) of the form (17) is equal to $\lambda exp\{-\mu x\}$, $\mu=const$. For the output function of the form $r(x)=a(1-\chi_x)$, a=const>0, (7),

$$P_{st}(y) = P_0[\delta(y) + \mu(1-P_0)exp\{-(\mu-\lambda/a)y\}];$$

$$P_0 = 1 - \rho / a = lim_{t \to \infty} P(x = 0, t); \quad \rho = \lambda / \mu. \tag{75}$$

Let us consider the behavior of the maximum of the distribution function $f(x)$. This maximum, as in [1], will be identified with the macroscopic phase of the substance. For (75) we find that at the values of the parameter $\rho/a=1$ the system qualitatively changes the nature of its behavior, a phase transition occurs. In this case, the condition for the existence of stationary distributions (38) is violated, and at $\rho > a$ there are no stationary states in the system. For the output function $r(x)=bx,\ b>0,\ P_0=0$, (22),

$$f(y) = P_{st}(y) = (\mu)^{\lambda/b} y^{(\lambda/b)-1} exp\{-\mu y\} / \Gamma(\lambda / b), \tag{76}$$

($\Gamma$ is the gamma function) the stationary probability of degeneration is equal to zero. The first phase transition occurs at those values of $\lambda/b,\ 0 \le \lambda/b < 1$, at which the normalization of the function (76) ceases to diverge (this occurs at $\lambda/b \ge 0$ or at $\mu > 0$, since at $\mu=0$, $P_{st}=0$ as well as at $\lambda/b=0$). The second phase transition is at $\lambda=b$. Similar results were obtained in Ref. [1] using Gaussian noise, where it is noted that the second transition does not have a deterministic twin, its existence is due exclusively to external noise. This feature of the stationary probability is interpreted as a nonequilibrium phase transition caused by external fluctuations.

The phase transition at $\lambda=b$ corresponds to the appearance of a non-zero maximum of the distribution. The divergence of the distribution at zero disappears, and the stationary distribution density looks like its peak nears the deterministic value (Fig. 1). If we choose the input in the form $b(x)=c^2 x exp\{-cx\},\ c=const$, then $P_{st}(y)=exp\{-\lambda/b\}(-\lambda/b)^{(1-\lambda/b)/2} y^{(\lambda/b-1)/2} J_{(\lambda/b)-1}[2(-\lambda y/b)^{1/2}]$ ($J$ is the Bessel function). One phase transition will also be at $\lambda=b$. For $r(x)=a(1-\chi_x)+bx$, (9) the stationary distribution density is equal to (18)-(19).

The value of $P_0^{-1}=1+(\mu a/b)^{(-\lambda/b)}(\lambda/b)exp\{-\mu a/b\}\Gamma(\lambda/b;\ \mu a/b)$ ($\Gamma(x;y)$ is the incomplete gamma function) is found from the normalization condition. One phase transition occurs at those values $\lambda/b$ and $\mu a/b$ of the arguments of the incomplete gamma function at which it begins to converge (at $\lambda>0$, as in (76), and at $b>0$, as well as at $\mu a/b<\infty$), - in this case the distribution leaves zero, but does not yet have a peak. The second one - at $\lambda=a\mu+b$. At $b\to 0$, these critical values of the parameters coincide with the results obtained for the output model $r=a$, and at $a\to 0$ - with the results of model (76).

As in Ref. [1], we can introduce the stochastic potential $U$, (78) rewriting $f(y)$ from $P_{st}(y)=P(y=0)+f(y)$ in the form $f(y)=exp\{-U(y)\}$. We find that $U(y)=(\mu-\lambda/a)y-ln(\mu P_0(1-P_0))$ for expression (75), $U(y)=\mu y-(\lambda/b-1)lny-\lambda(ln\mu)/b+ln\Gamma(\lambda/b)$ for expression (76), and $U(y)=\mu y-(\lambda/b-1)\ ln(a+by)+\lambda\ (lna)/b - ln\lambda - lnP_0$ for expression (18).

The above relations are written for the case $\beta=1$. The family of stochastic operators $\Phi_\beta$, depending on $\beta$, which is not macroscopic in nature, is related to the kinetic operator by the relation $\Phi_\beta(y, x) = \beta\ V(y/\beta, x)$ (31). The canonical Gibbs ensemble is defined for $\beta=(k_B T)^{-1}$, the value of $\beta$ takes into account the influence of the environment. In the general case, $\beta^{-1}$ is the ratio of the micro- and macroscopic scales. For the kinetic potential of the storage process, the component $\varphi_\beta$ depends on $\beta$; here $\lambda_\beta=\lambda\beta$; $b_\beta(x)=\beta b(\beta x)$; $\varphi_\beta(\theta)=\beta\varphi(\theta/\beta)$ (32). For the distribution (17) $\lambda_\beta=\beta\lambda;\ \mu_\beta=\beta\mu$. Then the relation for $U$ is replaced by the formula $f_{st}(y)=exp\{-\beta U_\beta(y)\}$, (78) and the expressions for $U(y)$ are replaced by the expressions $U_\beta(y)=(\mu-\lambda/a)y-\beta^{-1}ln[\beta\mu P_{0\beta}(1-P_{0\beta})]$; $U_\beta(y)=\mu y-\beta^{-1}(\lambda\beta/b-1)lny - \lambda ln(\beta\mu)/b+\beta^{-1}ln\Gamma(\lambda\beta/b)$; $U_\beta(y)=\mu y-\beta^{-1}(\lambda\beta/b-1)ln(a+by)+\lambda(lna)/b-\beta^{-1}ln\lambda\beta-\beta^{-1}lnP_{0\beta}$. The same result can be obtained from the solution of stationary kinetic equations with the value of $\varphi_\beta(\theta)$.

The above consideration allows us to suggest that storage models provide broader possibilities for describing phase transitions than the diffusion models traditionally used for this purpose Ref. [1]. One of the reasons for this may be the fact that the "basic" distributions that arise as a first approximation of the theory in the storage model are not Gaussian, but exponential and gamma distributions. Considering the importance of a detailed description of the behavior of physical systems in the vicinity of a phase transition (one can point, for example, to the behavior of a nuclear reactor, for which the vicinity of a phase transition represents a stationary operating mode), a detailed analysis of various aspects of the physical interpretation of storage models and their physical applications is essential. Real systems are apparently described by more complex output and input functions than the model expressions considered above. Above, we considered the stationary behavior of fluctuations, which determines the time scale on a set of instabilities. In the general case, the spectrum of the stochastic evolution operator should be analyzed.

### 8. Quasi-potential, external forces and evolution of the system

In Ref. [43] the quasi-potential of a dynamic system under the action of random disturbances is defined. In our case, the quasi-potential *U(A)* is equal to:

$$U(A)=\beta\int x(A)dA\,, \qquad f_{st}(y)=exp\{-\beta\, U_{\beta}(y)\}, \tag{78}$$

where the function $x(A)$ is determined from the relation (44). The same definition is used in [36], where the function $U_{\beta}(y)$ is called quasi-free energy. For (11) and (41), (38) the relation (44) takes the explicit form:

$$F(e^{\beta x})=e^{-a\beta x/b}(1-x)^{-\lambda\beta/b}[1-\frac{\gamma(\frac{\lambda\beta}{b}+1,\frac{a\beta}{b}(1-x))-\gamma(\frac{\lambda\beta}{b}+1,\frac{a\beta}{b})}{\Gamma(\frac{\lambda\beta}{b}+1)-\gamma(\frac{\lambda\beta}{b}+1,\frac{a\beta}{b})}]\,;$$

$$\beta A(x)=\frac{\beta}{b}[-a+\frac{\lambda}{1-x}+a\frac{[\frac{a\beta}{b}(1-x)]^{\lambda\beta/b}\exp\{-\frac{a\beta}{b}(1-x)\}}{\Gamma(\frac{\lambda\beta}{b}+1)-\gamma(\frac{\lambda\beta}{b}+1,\frac{a\beta}{b}(1-x))}], \tag{79}$$

where $\Gamma(\alpha)$ is the gamma function, $\gamma(\alpha,x)$ is the incomplete gamma function Ref. [40]

If, as in [1], we describe phase transitions by expressions

$$\frac{\partial U(A)}{\partial A}_{|A_c}=0, \quad \frac{\partial^2 U(A)}{\partial A^2}_{|A_c}=0\,. \tag{80}$$

then from (78) we obtain that the phase transition point can be sought from the expressions:

$$\beta x(A_c)=0, \quad \beta\frac{\partial x(A)}{\partial A}_{|A_c}=0\,. \tag{81}$$

When solving relations (79)–(81), various approximations are used, based on the results of [40], [44]. In the next section this approach is applied to the description of the behavior of micelles. In this case, the quantity $A(x)$ in (79) is the average value of a random internal thermodynamic parameter under external forces *x*, close to the stationary value. For example, the quantity $x\sim c_1\mu^0{}_1$, where $c_1$ is the concentration of single molecules in the solution, $\mu^0{}_1$ is the chemical potential of a single molecule in the solution, can act as thermodynamic forces *x* (or the force $h$ ratios $x-h$ associated with them Ref. [36]), The quantity $\beta$ in this case is proportional to $\beta = 1/k_B T$. Relation (79) expresses the influence of external forces on the average $A(x)$.

The function $f(A)$ from (45) is written from the expression for $\kappa_1(x)$, obtained from (36), after substituting there $x(A)$. In the stationary case $f(A_{st}) = 0$, the function $f(A)$ can be expanded in a series about the values $A_{st}$, it can also be expanded in a series in powers of $\beta^{-1}$. The solutions of the above equations yield two branches $x(A)$ and $f(A)$. One branch is stable, the second is unstable. For the stable branch, the potential has a minimum at the point $A_{st}$. The potentials corresponding to the values of the other branch do not have stationary stable values. When $\beta \to \infty$ we obtain a low-temperature (or low-noise) approximation of the form $dA/dt = \lambda - r(A)$. The quantity *A* tends to the stationary value (14) with increasing time.

For *ρ/a* > 1, there are no stationary states in the system and, consequently, no stationary structures *q(t)*. Let us assume that it is possible to form a new structural hierarchical level *q`(t)* associated with *q(t)*, for example, by the transformation *q`(t)=q(t)+c*. Then, in the stochastic equation for *q* and in the expression for the constant rate of exit of the form $r(q)=a(1-\chi_q)$, the term $\chi_q$ is replaced by $\chi_{q`-c}$.

This change can be attributed to *dA/dt* by writing *dA`/dt= ρ`= ρ-am*, where the value of *m* is related to the value of *c*. A situation is possible when *ρ`<a* at *ρ>a*. Then the structural level *q`* has a steady state *A`=A+c*. In this case, *ρ* does not change. Then *x(A`-c)* has a real minimum at $A`_{st}=c+A_{st}$, where $A_{st} = (\sigma^2/2a)/\beta(1-\rho/a) < 0$. If *A`*>0, then the dependence $U^-(A)$ at *ρ/a* >2 with a negative potential extremum passes into a dependence with a positive potential extremum, the unstable branch $x^-$, $U^-(A)^-$ for the changed hierarchical structure becomes a stable steady-state branch. A structural phase transition occurs in the system. For the dependencies *x`*, *U`* on *A`* at *ρ`*, expressions (51), (68) are valid.

From expression (68), written for model (7), we obtain from (78) that:

$$\beta^{-1}U^{\pm}(A) = (\mu - \frac{\lambda}{2a})A \mp \frac{1}{2}\frac{\lambda}{a}[A^{1/2}(A+d)^{1/2} + d\ln((A/d)^{1/2} + (1+A/d)^{1/2}], \quad d = \frac{4a}{\lambda\beta}. \quad (82)$$

Let us consider the behavior of the system depending on the noise intensity $\rho = \lambda\langle x\rangle = \lambda/\mu$ for (17). From (78):

$$x^+(A) = \frac{\partial(\beta^{-1}U^+(A))}{\partial A},$$

where $U^+(A)$ is written in (82). The equality of this value to zero, when $(\rho/2a)^{-1} - 1 = (1-d/A)^{1/2}$, corresponds to the extremum $U^+(A)$. The right side is positive when $1-\rho/2a > 0, \quad \rho/a < 2$. Then the extremum point coincides with the stationary value $A_{extr} = (\sigma^2/2a)/\beta(1-\rho/a) = A_{st}$. At the same time $U^+(A_{st}) = \ln P_0 = -\ln Q$, and $x^+$ corresponds to the positive branch. For the negative branch of expression (82) there is no extremum. When $1 < \rho/a < 2$ the value $A_{extr} < 0$ for the positive branch. At $\rho/a > 2$ the value $x^+(A)<0$, the value $x^-(A)$ may have stationary non-zero values $A_{extr}$, but they are negative. At *ρ=0*, $U^{\pm}(A) = \beta\mu A$. At $\rho \to \infty$, $U^+(A) \to -\lambda A/a, U^-(A) = const$. At $\rho/a > 1$ there are no stationary states in the system and, therefore, no stationary structures *q(t)*.

For the output function of the form (7) and the Laplace transform of the stationary probability density of the random variable q, expressions (8) and (9) are obtained, which are valid for all values of *ρ* and for any models of the function *φ*. They correspond to the low-noise approximation, which for the model (7) is written in (69) under the assumption $\beta^{-1} \to 0$. In this case, there is no dependence on $\chi_q$, $r(0+)=0$, the probability of degeneration $P_0=P_0(e^{-v})=0$ for any effects $v=x\beta$. The average lifetime of the system in this case tends to infinity. From (8) with (27) we obtain (9); $U(A)=\beta\mu[A-\rho\ln(A)/b]$. With $\rho<b; U(A_{st})>0, A_{st}<1$. With $\rho\geq b; U(A_{st})\leq 0, A_{st}\geq 1$.

The exit model (9) combines models Eqs. (7) and (22). The constant exit rate leads to a singularity at zero, a non-zero probability of degeneration, and the presence of several branches of the system's evolution.

### 9. Example: Application of storage theory to the formation of micelles

In Ref. [27], molecular aggregate molecules of surfactants, the micelle is regarded as a storage system stochastic theory of storage. It comes with a given distribution of the molecule and by the given law and exited. The processes of micelle formation are reduced to the aggregation (association, clustering) of molecules or ions. Such processes are described by storage models.

The main assumption made is that the micelle evolution and model (1) are supposed to be adequately matched. Indeed, monomer molecules enter and exit a molecular aggregate (micelle), which is the essence of model (1). One micelle is considered a stochastic storage system. Let us compare equation (1) with the kinetic equation of micelle formation, which is written in [45] as:

$$\frac{\partial c_n}{\partial t}=J_{n-1}-J_n, \quad J_n=j^+{}_n c_n-j^-{}_{n+1}c_{n+1} \quad (n=1,2,...),$$

where $J_n$ is the flow of aggregates in the size space Ref. [45], $c_n$ is the concentration of molecular aggregates $\{n\}$ containing *n* monomer molecules, $j^+{}_n$ is the number of monomers adsorbed onto the aggregate $\{n\}$ from the solution per unit time, $j^-{}_{n+1}$ is the number of monomers leaving the aggregate $\{n+1\}$ into the solution per unit time, $j^+{}_n>0, \quad j^-{}_{n-1}>0$. It is assumed that the number of molecules in the aggregate changes as a result of the absorption of monomers by the aggregate or their release from the aggregate. Then:

$$\frac{\partial c_n}{\partial t}=j^+{}_{n-1}c_{n-1}+j^-{}_{n+1}c_{n+1}-(j^-{}_n+j^+{}_n)c_n.$$

This equation coincides with equation (1) when $X(t) = Z=c_n$, $r[Z]=a+bZ$, $dA/dt=a+j^+{}_{n-1}c_{n-1}+j^-{}_{n+1}c_{n+1}$, $b=j^-{}_n+j^+{}_n$, and $a$ is a parameter corresponding to a constant, independent of $c_n$, yield of monomers from the aggregate $\{n\}$ per unit time. Below, a similar but slightly different approach is used, when the stochastic storage system is considered not as the concentration of aggregates $\{n\}$, but as one micelle. Then, from the relationships written in [45], we obtain a coincidence with equation (1) when $X(t)=Z(t)=n(t)$,

$$r[n]=a+bn, \quad \frac{dA}{dt}=j^+{}_S\frac{2n_S}{(\Delta n_S)^2}+j_n^+\frac{\partial c_n}{\partial n}, \quad a=j_S^+\frac{\partial c_n}{\partial n}, \quad b=j_S^+\frac{2}{(\Delta n_S)^2}\ ; \quad j_c^+=j_n^+{}_{|n=n_c}, \qquad j_S^+=j_n^+{}_{|n=n_S},$$

the meaning of the notations $n_S, n_c, \Delta n_S$ is given below. The output model is defined as (9), and the input distribution density is defined as $g(x)=\mu e^{-\mu x}$, when the function $\varphi(\theta)$ (2) is equal to $\varphi(\theta)=\lambda\theta/(\mu+\theta)$ (17). As, in Refs. [46, 47], we assume that the inputs into the micelle are single surfactant molecules. Then μ=1, since in (2) $1=\bar{x}=\mu^{-1}=\int_0^\infty xg(x)dx$, and $\partial^n[\varphi(-x)/(-x)]/\partial x^n{}_{|x=0}=n!\lambda$.

In general, the equation for the Laplace transforms $F(e^{\beta x},t)=\int_0^{\infty} e^{\beta xq} p(q,t)dq$ of the probability density $p(q,t)$ that at time *t* there are *q* particles, surfactant molecules (surface-active substances) in the system (one micelle), has the form (3), (10). We use the ratios (28)-(32).

In order for the dimensionality relations to be satisfied, we multiply the quantity $\beta$ by the energy parameter $\Delta E_0$ (equal, for example, to $G_{n_S}$), similar to how in [46] the work $W_n$ (Fig. 1) is expressed in units of $k_B T$: $W_n=(G_n-n\mu_1)/k_B T$, where $G_n$ is the chemical potential (Gibbs energy) of the micelle, $n$ is the number of molecules in the micelle, $\mu_1$ is the chemical potential of the monomer molecule. In the equilibrium canonical Gibbs ensemble, the dimensional parameter β will be replaced by the dimensionless parameter: $\beta=1/k_B T \to \Delta E_0/k_B T$, where $k_B=1{,}38\circ 10^{-23}$ is the Boltzmann constant, *T* is the absolute temperature. The Boltzmann constant is a small quantity; this is used in Ref. [36]. In the general case, the quantity $\beta$ takes into account the influence of the environment. We consider $\beta^{-1}$ it a small parameter.

The parameter *a* of the average exit velocity can be specified as:

$$a=n_S/\tau_{mic}, \qquad (83)$$

where $\tau_{mic}=\pi\Delta n_S\Delta n_c\exp\{W_c-W_S\}/j_c^{\ +}$ is the average lifetime of a micelle Ref. [48], $n_s$ is the aggregation number of a stable molecular aggregate, the stationary number of surfactant molecules in a micelle, at the bottom of the potential well, Fig. 2, at the maximum of the distribution *f(n)*, Fig. 3, $W_c$ and $W_s$ are the values of the work of formation of a surfactant molecular aggregate at the critical point $n_c$ of transition through the barrier and at the stationary point $n_s$, the height of the potential hump and the depth of the potential well of the work of formation of a surfactant molecular aggregate, respectively, are Fig. 2, the parameters $\Delta n_c$ and $\Delta n_S$ describe the half-widths of the potential hump and the potential well of the aggregation work $W_n$ Ref. [46], $j_c^+=j^+{}_{n|n=n_c}$ and $j_s^+=j^+{}_{n|n=n_s}$ are the numbers of surfactant molecules added to the aggregate from $n=n_c$ and $n=n_s$ molecules at points $n=n_c$ and $n=n_s$ per unit time. We consider only one potential well, the presence of only spherical micelles in the solution, and the case of a non-ionic colloidal surfactant.

The parameter $\lambda$, the intensity of entry into one micelle of $n$ molecules with radius *R*, is defined, as in Ref. [48], in the form:

$$\lambda=j^+{}_n=4\pi Dc_1\frac{vR^2}{D+vR^2}, \qquad \lambda_c=j^+{}_c=4\pi Dc_1\frac{vR_c^2}{D+vR_c^2}, \qquad \lambda_S=j^+{}_S=4\pi Dc_1\frac{vR_S^2}{D+vR_S^2}, \qquad (84)$$

where $\lambda_c$ and $\lambda_S$ are the numbers of monomers coming from the solution into the micelle from $n$ monomers per unit time for the critical and stationary micelle, *D* is the diffusion coefficient of monomers, $R_c$ is the radius of the micelle at the point $n=n_S$, $R_c$ is the radius of the micelle at the point $n=n_c$, the parameter *v* characterizes the rate of absorption of monomers by the micelle from the solution, $c_1$ is the concentration of monomers in the solution. The distribution *f(n)* in Fig. 3 describes the number of micelles containing *n* molecules, and not one micelle of *n* molecules. But these distributions are close to each other. It is possible to limit ourselves to specifying only one parameter, for example *λ* (84), expressing the parameter *a* through it.

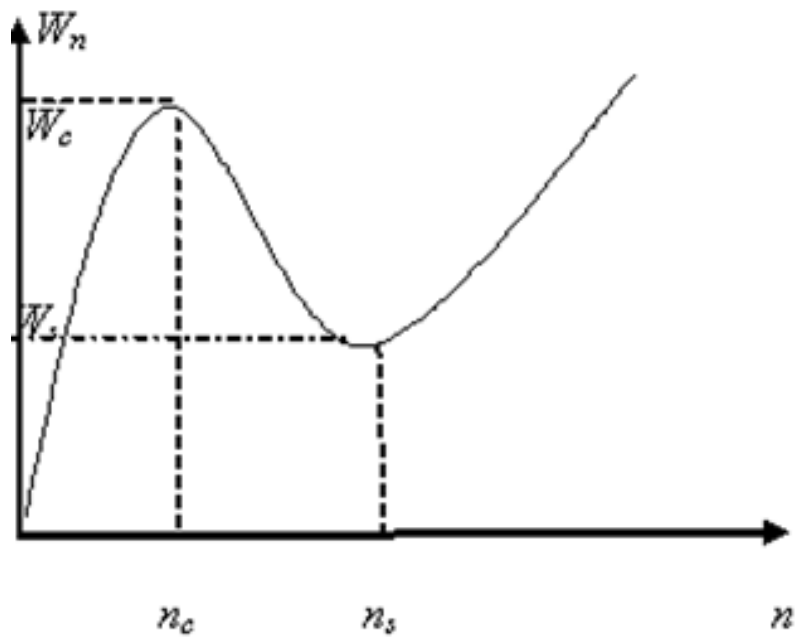

**Fig.2**. Behavior of the work of formation $W_n$ of a molecular aggregate of a surfactant depending on the aggregation number n at the existing potential well.

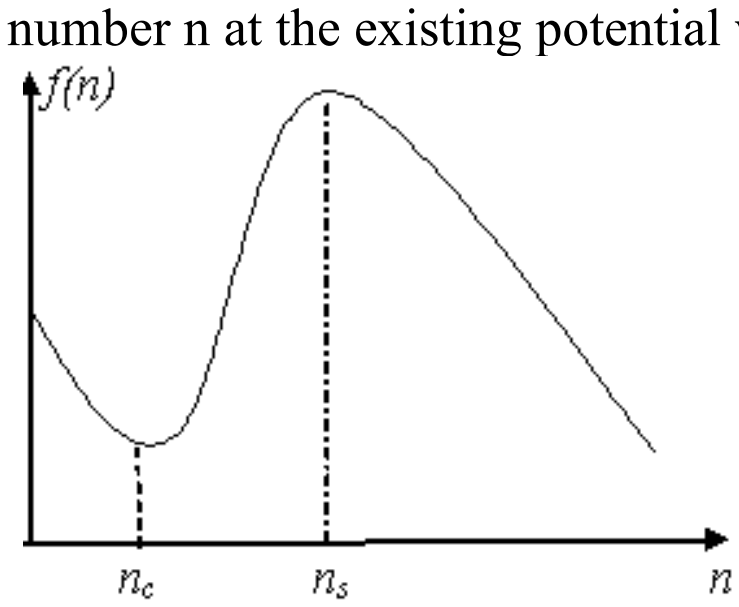

**Fig.3**. Behavior of quasi-equilibrium stationary distribution density f(n) for the number of surfactant molecules n in one micelle, corresponding to Fig. 2.

Figure 3 shows the distribution *f(n)* corresponding to the potential of Figure 2, since according to the Boltzmann distribution the distribution of the number of aggregates of *n* molecules is proportional to $e^{-W_n}$. But the stationary distribution of the storage process with output (9), the stationary solution of equation (10) has the form (11), (18), (19):

$$w_{st}(x) = P_0[\delta(x) + f_{st}(x)], \quad f_{st}(x) = e^{-\beta U_\beta(x)}, \tag{85}$$

$$U_\beta(x) = x - \frac{1}{\beta}\left(\frac{\lambda\beta}{b} - 1\right)\ln(a+bx) + \frac{\lambda \ln a}{b} - \frac{1}{\beta}\ln(\lambda\beta) - \frac{1}{\beta}\ln P_{0\beta}, \quad P_{0\beta}^{-1} = 1 + \left(\frac{\beta a}{b}\right)^{-\frac{\beta\lambda}{b}} \frac{\beta\lambda}{b} e^{\frac{\beta a}{b}} \Gamma\left(\frac{\beta\lambda}{b}, \frac{\beta a}{b}\right),$$

where $\Gamma(a,x)$ is the incomplete gamma function Ref. [40]. The continuous part of the distribution (85) has the form of a gamma distribution, Fig. 1.

The dependencies of the type in Fig. 3 can be obtained with a different type of output function. Thus, in Ref. [4] for the output function $r(q)=bq-cq^2(1-q)$ ($c, b \geq 0$), corresponding to the nonlinear volt-ampere characteristic, an expression for the stationary distribution density is written, giving a picture similar to Fig. 3. However, we strive to simplify the description as much as possible. Therefore, we will proceed from the model (9), (17), but assume that the range of values is considered. In Fig. 2 the origin of coordinates corresponds to the value $n=n_c$, we consider the function $f(n-n_c)$. A corresponding shift occurs along the vertical axis. As in Ref. [46], we will consider two regions: a near-critical one, in the region of $n_c$ values, and a micellar one, in the region of $n_s$. In [45], the regions of the potential hump and the potential well are specified by the relations $n_c - \Delta n_c \leq n \leq n_c + \Delta n_c$, $n_S - \Delta n_S \leq n \leq n_S + \Delta n_S$, respectively.

By equating the truncated distribution (85) to the truncated distribution modeling the behavior of the storage process distribution for $n = n_S$ (normal distribution) and $n = n_c$ (linearly decreasing distribution), we determine the mathematical parameters of the storage model (*λ, a, b*, one of them is given by the relation (83) or (84) through the physical parameters of the problem. For the micellar region near $n = n_S$ we obtain that for a given value of λ (84):

$$b_S = \frac{\lambda_S}{\frac{\beta}{2}\Delta n^2{}_S + \frac{2}{\beta}}; \qquad a_S = \lambda_S \frac{\frac{\beta}{2}\Delta n^2{}_S - n_S}{\frac{\beta}{2}\Delta n^2{}_S + \frac{2}{\beta}}; \qquad \frac{\lambda_S - a_S}{b_S} = n_S + \frac{2}{\beta}. \tag{86}$$

For the near-critical region near $n = n_c$, $n_c - \Delta n_c \le n \le n_c$, we find that:

$$b_c = \frac{\lambda_c \beta}{1 + (1 - \beta m \Delta n_c)^2}; \qquad a_c = \lambda_c \beta \frac{\beta (m \Delta n_c)^2 - n_c}{1 + (1 - \beta m \Delta n_c)^2}; \qquad \frac{\lambda_c - a_c}{b_c} = \frac{1}{\beta}[2 + \beta(n_c - 2m\Delta n_c)]; \quad m \ge 0, \tag{87}$$

where in the truncated distribution (85) of the form $\omega_{tr\,st} = C_{tr} e^{-\beta x}(a + bx)^{\frac{\lambda\beta}{b} - 1}$, $C_{tr} = b e^{-\frac{\beta a}{b}} (\beta / b)^{-\frac{\beta\lambda}{b}} / \Delta\gamma$, $\Delta\gamma = \gamma(\beta\lambda / b, (\beta / b)((c_2 - a) / b)) - \gamma(\beta\lambda / b, (\beta / b)((c_1 - a) / b))$ we set the values of the argument *x* in the form $x = n_c \pm m\Delta n_c$. The conditional truncated distribution of a random variable $\xi$, provided that its values lie in the region $c_1 \le \xi \le c_2$, is written in the form:

$$F(x | c_1 \le \xi \le c_2) = \begin{cases} 0 & x < c_1 \\ \frac{F_\xi(x) - F_\xi(c_1)}{F_\xi(c_2) - F_\xi(c_1)}, & c_1 \le x \le c_2 \\ 1, & x > c_2 \end{cases}. \tag{88}$$

The truncation for expression (85) occurs in the interval $n_S - \Delta n_S \le n \le n_S + \Delta n_S$, and for (88) $c_1 = n_c$, $c_2 = n_c \pm m\Delta n_c$. Thus, the expression for the parameter λ is given from Ref. [48], the parameters a, b are determined from a comparison of the model distribution (85) with other model distributions approximating the behavior of the functions in Fig. 2, 3 in the neighborhood of the points $n = n_S$ and $n = n_c$.

In this case, both the type of distribution and the values of the parameters in the micellar stationary region and in the critical region differ, the entry and exit processes occur with different intensities. It is assumed that such stationary parameters as $\langle n \rangle = n_S$, $n_c$, $\Delta n_S$, $\Delta n_c$, $\lambda$ are known from [45-48]. From (86)-(87) it is evident that the parameter *b* is much smaller than *a* and *λ*, since $\beta^{-1}$<<1. For the values of the parameters given in Refs. [45-48] for the droplet model, from (86) we obtain that in the micellar region $b_S \sim \lambda_S / 562$, $a_S \sim \lambda_S / 1.1$. For the critical region, from (87) we find $b_c \sim \lambda_c / 706$, $a_c \sim 0{,}99\lambda_S$; $j^+{}_c \sim 5$, $j^+{}_S \sim 8$. It is evident that, although the analytical expressions (86) and (87) and the values of *λ* differ for these regions, the values of the parameters *a* and *b* differ insignificantly.

Using the recorded relationships in [27], the non-stationary behavior and equilibrium state of micelles are described, the conditions for the formation of micelles and the lifetime of micelles are considered in detail. In addition to the characteristics of micelles considered in [27], other parameters can be determined. External effects can be taken into account using the methods of Ref. [36]. The condition of existence of stationary states (38) and lifetime of micelles plays a significant role in the description of physicochemical systems similar to micellar ones.

## 10. Another approach to the storage model

Another approach to the stochastic storage model was proposed in Ref. [49]. This approach has a direct physical meaning and shows the generality and importance of stochastic storage theory in a wide variety of applications.

Some difficulties in the study of processes (1) are due to the non-uniqueness of the solution of the equation dy(t)=-r(y(t))dt and are not related to the probabilistic structure of X(t). Another definition of a wider class of storage processes, free from the indicated drawback, is proposed in Ref. [49]. It is based on the observation that processes of type (1) fit into the general scheme of dynamic systems subject to the additive effect of process A(t). This definition and some properties of the storage process are given in [49], and a truncated storage process is considered, taking values from a certain compact set [0, a], and the average times for reaching the lower level by inventory storage processes are investigated. Conditions for the existence of stationary distributions, different from (38), and equations for distributions, different from (40), are obtained.

The approach proposed in Ref. [49] extends the applicability of storage processes. It directly links storage processes to dynamic systems and replaces the most commonly used random disturbances in the form of white noise with more generalized Poisson processes.

In Ref. [49], the quantities:

$$\tau(y)=\inf(t: X(t)=y),\quad \tau(x, y)=\inf(t: X_x(t)=y),\quad m(x, y)=\boldsymbol{E}_x\tau(y)=\mathbf{E}\tau(x, y),$$

are defined where $X_x(t)$ is the Markov process defined in [49], in which the output function r(x) in (1) is replaced by a "dynamic system" $g_x(t)$ with the properties specified in [49]. The quantities:

$$\theta(y, a)=\inf(t: Y(t; a)=y),\quad \theta(x, y; a)=\inf(t: Y_x(t; a)=y),\quad t(y, x)=\inf(t: g_y(t)=x)\ (y>x),$$
$$n(x, y; a)=\mathbf{E}_x\theta(y; a)=\mathbf{E}\theta(x, y; a),$$

are defined where Y(t;a) is a truncated storage process. In Ref. [49] theorem 2 was proved: If $\lambda<\infty$ and the condition $t(x,y)<\infty$ is satisfied for $x\geq y>0$, then n(x,y;a) for $a\geq x\geq y\geq c>0$ is the unique measurable bounded solution of the equation:

$$n^*(x, y; a) = t(x, y) + \int_y^x \int_0^\infty [n^*((u+z)\wedge a, y; a) - n^*(u, y; a)]\nu(dz)d_u t(u, y)\,. \qquad (89)$$

Since $x\geq y>0$, then Theorem 2 on the moment of reaching the lower (relative to the initial value x) level. The same equation (89) is also true for the function m(x, y). In Ref. [49] Theorem 4 was also proved: for the existence of a stationary distribution Π of a process X it is necessary that for some y and all x>y:

$$m(x,y)=\sum_{n=0}^{\infty} K^n\, t\{x, y)<\infty,\quad x>y,\quad Kf(x) = \int_y^x \int_0^\infty [f(u+z) - f(u)]\nu(dz)d_u t(u, y)\,. \quad (90)$$

In Ref. [50] equations (26), (39), (40) were obtained. The results of this article generalize the statements proved in the work [9] for processes satisfying equation (1). In Ref. [51] an equation was obtained for $\varphi(s; x, y) = \boldsymbol{E}_x e^{-s\tau(y)}$, it was proved that if $\lambda<\infty$, then for s>0 the function φ(s;x,y) is the only bounded measurable in x solution of the equation:

$$\varphi(s; x, y) = e^{-(\lambda+s)t(x,y)} + \int_y^x e^{-(\lambda+s)t(x,u)} \int_0^\infty \varphi(s; u+r, y)\nu(dr)d_u t(u, y)\,. \qquad (91)$$

It is possible to solve equation (91) explicitly in very rare cases. One of them is when in (1) r(x)=μx, and $\nu(y,\infty)=\lambda e^{-\nu y}$, λ>0, μ>0, ν>0. However, solutions of linear equations of the form dX(t)=- μX(t)dt+dA(t) describe a whole series of real processes. If $\nu(y,\infty)=\lambda e^{-\nu y}$ for y>0, then:

$$\varphi(s; x, y) = F(\frac{s}{\mu}, \frac{s+\lambda}{\mu}+1, x\nu)\int_{x\nu}^\infty \frac{dz}{F^2(\frac{s}{\mu}, \frac{s+\lambda}{\mu}+1, z)p(z)} \Big/ F(\frac{s}{\mu}, \frac{s+\lambda}{\mu}+1, y\nu)\int_{y\nu}^\infty \frac{dz}{F^2(\frac{s}{\mu}, \frac{s+\lambda}{\mu}+1, z)p(z)},$$

$$p(x) = e^{-x} x^{\frac{s+\lambda}{\mu}+1}.$$

Here F(a,b,x) is the Pochhammer function, which has the representation Ref. [52]:

$$F(a,b,x) = 1 + \sum_{k=1}^{\infty} \frac{a(a+1)...(a+k-1)}{b(b+1)...(b+k-1)} \frac{x^k}{k!}.$$

In Ref. [53] a classification of the behavior of storage processes was carried out for another case, when in (1) the power function of the output r(x)= $x^{\beta}$, $\beta \geq 0$, stable input, when:

$$\boldsymbol{E}\exp[-sA(t)] = \exp[-t\int_0^{\infty}(1-e^{-su})\nu(du)], \quad \nu(du) = Au^{-1-\alpha}, \ 0<\alpha<1, \ A>0.$$

The conditions of recurrence, moments of reaching the zero level, and local time at zero were obtained.

In Ref. [54] it is shown that for **E**A(1)<∞ the criterion for the existence of a stationary distribution is the condition $\lim_{x\to\infty}$r(x)>**E**A(1). In Ref. [54], as in Ref. [49], dynamic systems are considered with the construction of the corresponding process, called there a process with deterministic drift. Note that the transition from dynamic deterministic systems to stochastic behavior was also considered in Ref. [55] and in a number of other works, where K-systems and Kolmogorov flows are related to such systems. This is an approach from the side of dynamics, dynamic systems, and in Refs. [54], [49] – from the side of stochastic.

In [54] a lemma is also proved: there exist output functions r(x) such that the associated inventory storage process has a stationary distribution for an arbitrary input process A(t).

Equation (91) in [56] for $\varphi(s;x,y) = \boldsymbol{E}_x e^{-s\tau(y)}$ is written as:

$$\ln\varphi(s;x,y) = -st(x,y) + \int_y^x\int_0^{\infty}(\varphi(s;u+v,u)-1)\nu(dv)d_u t(u,y), \quad \text{t(x,y)=}\int_y^x \frac{du}{r(u)}.$$

**11. Other examples**

**11a. Possibilities of using stochastic storage theory in tribology problems**

B Ref. [57] methods of statistical physics and probability theory are applied to tribological problems. First-passage time (FPT) methods, effective in application to problems of the time to failure or the end of a certain stage of a process, were applied to tribological studies in Ref. [57].

In Ref. [57], the controlled parameter of the state of the tribological system $X_t$, which changes with wear, the linear size of one of the parts or their combination, is assumed to be a random variable. In the linear wear model, the state parameter $X_t$ changes as:

$$X_t = X_0 - \Gamma_y t, \tag{92}$$

where $X_0$ is random variable of the state parameter $X_t$ at the initial time $t_0$, $\Gamma_y$ is a random variable linear wear rate with mean value $\gamma_y$. The limiting (normative) value of the state parameter $X_t$ is equal to $X_{lim}$. In Ref. [57], the parameter $\gamma_y$ is expressed in terms of the ratio of the nominal areas of contact and friction of the elements, the critical value of the measure of damage to materials, the speed of the relative movement of the surfaces, and the coefficients of conversion (absorption) of external energy by the surface layer of triboelements.

It is required to establish an adequate correspondence between a physical phenomenon and a random process used for its mathematical modeling. There are many random processes and the first-passage time (FPT) statistics that can be adapted to a specific task. The use of various types of approximations significantly depends on the stage of system evolution.

Let's apply possibilities of linking the FPT distribution with the thermodynamic characteristics of the system to the linear wear model (92). Consider reaching the limit value of the parameter $X_t$ (92) $X_{lim}$ as a random process reaching a certain level. As an independent thermodynamic parameter, a random time of the first achievement of the set value $X_t = X_{lim}$ is chosen.

The behavior of the collapsing segment $X_0$-$X_t$ from expression (92) in Ref. [57] is modeled by the normal distribution. Let us model it with a random process of the form:

$$\frac{dv}{dt} = \mu + \xi(t)\,, \tag{93}$$

where $v=X_0-X_t$ from (92) is the length of the section under consideration, μ is the rate of its destruction; in accordance with (92) $\mu = \gamma_y$ from (92), $\xi(t)$ is normal white noise; $\langle \xi(t) \rangle = 0,\ \ \langle \xi(t_1)\xi(t_2) \rangle = \sigma^2 \delta(t_1 - t_2)$. Process (93) is considered in the interval (0, h), where at point 0, corresponding to point $v= X_0-X_t=0$ from (92), a reflecting screen is placed, and at point $h$, corresponding to point $X_{lim}$ from (92), an absorbing screen is placed.

In tribology problems, the action is usually impulsive. Is placed, instead of white noise, as in (93), a generalized Poisson process should be used, as in the sections above and in [4, 22].

### 11b. Modeling of coagulating aerosol systems using a storage model

In Ref. [17] a general stochastic approach to the description of the coagulating aerosol system is developed. As the object of description one can consider arbitrary mesoscopic values (number of aerosol clusters, their size, etc., and various boundary value problems Ref. [58]).

The stochastic description of the coagulation process was performed in papers like [12-17, 59-62]. So, the birth-and-death model leads to the kinetic equation of coagulation in the form:

$$\partial P(t,\ [X]) / \partial t \ = \ (1/2)\sum_{i \neq j} W(i,j)\ \left[ (X_i + 1)(X_j + 1) P(X_i + 1, X_j + 1, X_{i+j} - 1) \ - \ X_i X_j P \right] +$$
$$+ (1/2)\sum_i W(i,i)\ \left[ (X_i + 2)(X_i + 1) P(X_i + 2,\ X_{2i} - 1) \ - \ X_i\,(X_i - 1) P \right], \tag{94}$$

where P(t, $X_1, X_2, ..., X_n$,...) is the probability to find $X_i$ particles (clusters) having the size i (i=1, 2,...) in the time t; W*(i,j)* is the coagulation probability per time unit of the particles i and j (containing, in general, the factor $L^{-3}$, where L is the size of a system). The equation for the generating functional:

$$F(t,[s]) = \sum_{[X]} \Pi_i \,(s_i)^{Xi}\ P(t,[X]) \tag{95}$$

can be drawn from (94):

$$\partial F / \partial t = (1/2)\sum_{i,j} W(i,j)(s_{i+j} - s_i s_j)\, \partial^2 F / \partial s_i \partial s_j\ . \tag{96}$$

The equation for the average number of clusters (from either (94) or (96)) has the form:

$$\partial < X_k > / \partial t = (1/2)\sum_{ij} W(i,j) D(i,j \,|\, k) Q_2(i,j); \tag{97}$$

$$D(i,j \,|\, k) = \delta(i+j;k) - \delta(i;k) - \delta(j;k)\,; \quad Q_2(i,j) = \partial^2 F / \partial s_i \partial s_j \,|_{[s]=1} = < X_i (X_j - \delta(i;j)) > \tag{98}$$

(δ(i;j) is the Cronecker symbol) is unclosed since higher momenta $Q_k$ are involved for which successive set of equations can be derived from (94), (96). If one makes an assertion that the random number of clusters of each size has the independent Poisson statistics then:

$$Q_2(i,j) = < X_i (X_j - \delta(i;j)) > = < X_i > < X_j >, \tag{99}$$

and one arrives at the Smoluchowski equation from (97):

$$\partial < X_k > / \partial t = (1/2)\sum_{ij} W(i,j) D(i,j \,|\, k) < X_i > < X_j > . \tag{100}$$

In Ref. [63] the transition from (97) to (100) was performed based on the method of van Kampen [3]. In Refs. [62, 64] the spatially inhomogeneous coagulating systems were treated using the discretization operations both in space and time. The stochastic storage model for the random number of monomers in a cluster was introduced in Refs. [13, 65]. The general stochastic approach for describing arbitrary (random) macroscopic values characterizing an aerosol system is presented. The traditional stochastic storage model is generalized and the results are applied to the investigation of coagulating systems.

The mesoscopic level deals with the distribution function (or stochastic process) for the order parameters whose averages are to be treated macroscopically as thermodynamic quantities. For an aerosol system one could point out such values as number of clusters in a unit volume, size of a given cluster treating them as the order parameters. Denote such an order parameter as q(t) without its concretization for a moment (q(t) can be readily understood as multicomponent vector as well). In the assumption of the Markovian character of a process the distribution function ω(q,t) satisfies the master equation of the general type (Eq. (33) at p=ω, see, for example, Ref. [36]):

$$\partial\omega(q,t)/\partial t = N_{\partial,q}\,\Phi(-\partial/\partial q, q)\,\omega(q,t), \tag{101}$$

Thus, the fundamental quantity of the mesoscopic approach is the matrix of transition probabilities (or kinetic operator which is nothing more or less than the generating function on these probabilities). The Laplace transform of the function ω(q,t):

$$F(exp\{-\theta\},t) = \int_0^\infty exp\{-\theta y\}\omega_t\left(y\right)dy \tag{102}$$

resembles (95) up to the substitution s=exp{-θ}. The kinetic equation in this representation (that is for F(exp{-θ}) (from (102)) is (34):

$$\partial F(\exp\{-\theta\},t)/\partial t = N_{\theta,\partial/\partial\theta}\,\Phi(-\theta,-\partial/\partial\theta)\,F(\exp\{-\theta\},t). \tag{103}$$

We offer some common examples illustrating the specification of the transition matrix (kinetic operator). The diffusion process is by definition:

$$\Phi(\theta,q) = K_1(q)\theta + K_2(q)\theta^2/2, \tag{104}$$

where $K_1$, $K_2$ are drift and diffusion coefficients respectively. Substituting (104) into (101) yields the Fokker-Planck equation. This approximation is quite common in many areas of physics, and it was applied to various effects in the aerosol systems Ref. [66] - such as the spatial diffusion, filtration, coagulation, sedimentation processes. The value $K_1$ from (104) was thus $V_x$ - the projection of the velocity of the aerosol cluster onto axis x; $K_2/2=D=const$, D is the diffusion coefficient. As the random value q one took the cluster coordinate x (which may be called some external coordinate). The description of a single (separate) cluster was thus performed. A single cluster description was also introduced in Refs. [13, 65]. As the random number q we took the number m of monomers in a cluster, that is the internal coordinate. The stochastic storage model was used in the kinetic equation for this value:

$$dm/dt = dA/dt - r[m(t)], \tag{105}$$

where A(t) is a random input function, r[m] is the release rate. Equation (105) coincides with equation (1) when m=X(t), $r=r_\chi$. The input A(t) is given by specifying the cumulant function (2).

Thus, a number of monomers in a single arbitrary choice cluster is treated as random storage in a storage system. For the process (104)-(105):

$$\Phi(-\theta,m)=-\varphi(\theta)+\theta r_\chi(m);\quad r_\chi(m)=r(m)-r(0+)\chi_m;\quad \chi_m=1, \text{ if } m=0,\ \chi_m=0, \text{ if } m>0. \tag{106}$$

From (101), (106) obtain:

$$\partial\omega\left(m,t\right)/\partial t = \int_0^\infty (exp\{-y\partial/\partial m\}-1)\lambda_m b_m\left(y\right)\omega\left(m\right)dy + \partial(r_\chi\left(m\right)\omega\left(m\right))/\partial m\,. \tag{107}$$

Let's take a pure coagulation process. One traces the fate of an arbitrary chooses cluster supposing that it remains the same in all coagulations, even if it coagulates with larger clusters. Thus, the cluster can only grow and only the input term in (105) is present (that is r(m)=0). Now make a conjecture as to the shape of ω(m), λ, b(x). We assume:

$$\omega(m,t) = n(m,t)/N\,; \quad N = \int_0^\infty n(m,t)\,dm\,; \quad \lambda b(x) = \beta(m,x)\,n(x)/2\,, \qquad (108)$$

where n(m,t) = <$X_m$>/$L^3$ is the concentration of clusters with m monomers, β(m,x)=W(m,x)$L^3$ is the coagulation coefficient, that is the core of the kinetic coagulation equation. The factor 1/2 arises because one accounts for on coagulation act twice. Substituting (108) in (107) at r=0 gets:

$$(\partial n(m)/\partial t)/N - (\partial N/\partial t)\left(n(m)/N^2\right) = (1/2)\int_0^\infty [\beta(m-y,y)\,n(y)\,n(m-y)/N - \beta(m,y)\,n(y)\,n(m)/N]\,dy\,. \quad (109)$$

It is worthwhile mentioning that the choice of ω and λb(x) in (108) is not quite correct from the point of view of the traditional storage model. There is a dependence of λb(x) in (108) on ω(x) (through n(x)=ω(x)N), on t (through the time dependence in n(x,t), β[m(t), x(t)]) and on m (through β(m, x)) which is in contradiction to the primary suppositions of Ref. [66]. Nevertheless, we start from (107) supposing that its solution satisfies (108). It was this approximation that led to (109) which in its turn yields the Smoluchowski equation of the free coagulation:

$$\partial\omega(m)/\partial t = N[(1/2)\int_0^m \beta(m-y,y)\,\omega(m-y)\,\omega(y)\,dy - \int_0^\infty \beta(m,y)\,\omega(y)\,dy\,\omega(m)] - \omega(m)\,\partial lnN/\partial t\,, \quad (110)$$

if:

$$-\partial lnN/\partial t \cong (1/2)\int_0^\infty\int_0^\infty \beta(x,y)\,\omega(x)\,n(y)\,dxdy \cong (1/2)\int_0^\infty \beta(m,y)\,n(y)\,dy\,. \qquad (111)$$

First identity in (111) corresponds to Smoluchowski's equation and the second one arises from the fact that one chooses a cluster m at random: whatever cluster of the system can figurative instead.

Thus the Smoluchowski equation (110) is obtained from the general master-equation of the storage theory under the following assumptions: a) one supposes in (107) r=0; given cluster is constantly growing, b) the solution of (107) should satisfy (108) if r=0, the dependence of λb(x) on m creates obstacles to the use of the storage model because c) the cluster m is arbitrary and can be replaced by any other cluster of the system which implies (111).

If one considers, like in (94)-(100) a random number of clusters $X_i$ of the size i, the corresponding multicomponent kinetic potential takes on the form:

$$\Phi(v_1,v_2,...;X_1,X_2,...) = \sum_{ij}\left[\left(exp\{v_{ij}\}-1\right)/2\right]\,W(i,j)\,X_i(X_j-\delta(i,j))\,; \quad v_{ij} = v_{i+j} - v_i - v_j\,. \quad (112)$$

Substituting (112) into (101) leads directly to (1); into (10) with the transition from (100) to (95) yields (96). As random value q one can take either the cluster energy, velocity, charge etc. or several such values simultaneously. In Ref. [66] the backward Chapman equation for the model (104) was applied to a number of important problems in the theory of aerosols (such as the time of diffusive sedimentation etc.). This equation was used in Ref. [67] for investigating the lifetime of aerosols in the storage model. (Under the notion of lifetime, we understand the random time moment of the degeneration of a cluster). The degeneration of all clusters means, for example, the resolving of a cloud. Another outcome of the evolution can be the coagulation of all clusters into a big one which means the precipitation. One more possible general result of the evolution is the transition to the domain of states with no stationary Ref. [18]. The physical manifestation of these effects is either the destruction of a system or at least some phase transition. The storage model seems to fit better for elucidating these occurrences than the Smoluchowski equation (103), birth-and-death model (94), (112) or diffusion approximation (104).

Let us generalize the storage model (106). The transform of kinetic potential Φ(x,q) is written, according to Ref. [36] as (50):

$$R(y,x) = \int\Phi(y,q)\,\omega_x(q)\,dq = lim_{\tau\to 0}(\tau^{-1})\int\left[exp\{y(q_2-q_1)\}-1\right]\times \qquad (113)$$

$$\times exp\{-xq\}\,p(q_2\,|\,q_1)\,\omega_{st}(q_1)\,dq_1dq_2\,/\,F(x) = \sum_{n=0}^\infty y^n\kappa_n(x)/n!\,,$$

where $p(q_2|q_1)$ are transition probabilities for the Markovian process, $\omega_{st}(q)$ is the stationary distribution, $\omega_x(q)=\exp\{-xq\}\omega_{st}(q)/F(x)$ (43); $F(x)=\int\exp\{-xq\}\omega_{st}(q)dq$; $\kappa_n(x)=\int K_n(q)\omega_x(q)dq$; $K_n(q)=(\tau^{-1})\int(q_2-q_1)^n p(q_2|q_1)dq_2$ (46) are kinetic coefficients. The fluctuation-dissipation relations in nonequilibrium stationary cases take on the form:

$$R(x,x) = 0. \quad (114)$$

For the diffusion process (104) $\kappa_n=0$ at $n\geq3$, then $R_D(y,x)=y\kappa_1(x)[1-y/x]$ (as seen from (104) and (114)). For the storage scheme another approximation is used: namely, independence of $\kappa_n(x)$ on x, $n\geq2$ (or independence on q of $K_n$ , $n\geq2$). Then:

$$R_S(y,x) = y\kappa_1(x)\,[1-\kappa_1(y)/\kappa_1(x)]. \quad (115)$$

In Ref. [36] the diffusion schema was adopted as the basic one to which successive amendments were considered. We develop a similar extension procedure for the storage scheme assuming the following series for $K_n$:

$$K_n(q) = K_{n,0}+\gamma K_{n,1}\, q+\gamma^2 K_{n,2}\, q^2/2!+...+\gamma^k K_{n,k}\, q^k/k!+\dots, \quad (116)$$

where $\gamma$ is the formal expansion parameter. The kinetic potential takes on the form:

$$\Phi(y,q)=\sum_{n=1}^{\infty} y^n K_n(q)/n! = yK_1(q)\sum_{k=0}^{\infty}(\gamma^k q^k/k!)(\sum_{n=2}^{\infty} y^n K_{n,k}/n!) = yK_1(q)-\eta_o(y)- \quad (117)$$

$$-\gamma\eta_1(y)q-\,...\,-\gamma^k\eta_k(y)q^k/k!\;-...;\;-\eta_k(y)=\sum_{n=2}^{\infty}y^n K_{n,k}/n!;\quad \eta_0(y)=\varphi(-y)+y\rho;\;\rho=\partial\varphi(\theta)/\partial\theta\,|_{\theta=0},$$

and its transform is:

$$R(y,x)=y\kappa_1(x)-\eta_0(y)-\gamma\,\eta_1(y)<A(x)>-...-\gamma^k\,\eta_k(y)<A^k(x)>/k!-...;\quad <A^k(x)=\int q^k\omega_x(q)dq. \quad (118)$$

The expression (118) can be regarded as some series on the basis 1, $<A(x)>$, $<A^2(x)>$,...., which naturally arises from the shape of stationary distribution for the concrete case, that represents the "eigen" basis for the problem. The full series (118) is, of course, equivalent to the full series of Gaussian scheme, but usually we intend to truncate the series at some $<A^k(x)>$ ; this form (contrary to the usual method implying the truncating at the term $x^k$) seems to be more convenient, for example, for investigating chaotic systems because it is quite natural to investigate their characteristics (in particular, $K_n(q)$) as arising from some averaging procedure over the areas of parameter space which yields either constant or smoothly varying in q coefficient functions. Applying (114) to (118) gets:

$$R_S(y,x)=y\kappa_1(x)[1-\kappa_1(y)/\kappa_1(x)]-\sum_{k=1}^{\infty}\gamma^k\varphi_k(y)/k!\,\left[<A^k(x)>-<A^k(y)>\right] \quad (119)$$

with "arbitrary" $\varphi(x)$. The functions $\varphi$ are thus "dissipative undetermined" Ref. [36] in the macroscopic approach of (117)-(119) which is based on taking as primitive (initial) quantities the "observables", that is a) stationary distributions and b) equations of motions (stored, as one can easily check, in $\kappa$). Another approach to be model specification consists in of specification rather the than process generating these observables than the observables themselves. Thus, we arrive at the specification of the process in terms of jump input and release rates. Split $K_1(q)$ in (117) into two parts $K_1(q)=\rho-r(q)$, where r(q) is an arbitrary release function and $\rho(q)$ enters into (117) as the term with n=1. One can write the series analogous to (116): $\rho(q)=\rho_o+\gamma\,\rho_1\,q+\gamma^2\,\rho_2\,q^2/2!+...$, where $\rho_l$ are coefficients $K_{n,k}$ from (116), n=1, k=0,1,2,... In this case we arrive at the kinetic potential analogous to the ordinary storage model (106):

$$\Phi_S(y,q)=-yr(q)+\sum_{k=0}^{\infty}(\gamma^k q^k/k!)(\sum_{n=1}^{\infty}y^n K_{n,k}/n!). \quad (120)$$

Functions:

$$-\varphi_k(y)\;=\sum_{n=1}^{\infty}y^n K_{n,k}/n! \quad (121)$$

can be interpreted (like (95)) in terms of input functions. Generalized input intensity is $\lambda(q)=\lambda_o+\gamma\lambda_1 q+\gamma^2\lambda_2 q^2/2+...$; $\lambda_k=\varphi_k(y=-\infty)$; distribution unction $b(\Delta,q)$ is related to function $\varphi=\sum_k \varphi_k\gamma^k q^k/k!=\lambda-\lambda\int exp\{-y\Delta\}b(\Delta,q)d\Delta$.

For example, we can set all functions $\varphi_l(y)=\varphi(y)\lambda_l$ equal within a factor (proportional); this is the situation of the birth-and-death processes. This construction is to some extent analogous to that of Ref. [68] if one takes the modulating process I(t) [68] coinciding with the main process q(t).

Substituting (120) into (103) we obtain the relation:

$$\partial F(x,t)/\partial t=-xr(d/dx)F-\varphi_0(x)F-\sum_{k=1}^{\hbar}\gamma^k\varphi_k(x)(d/dx)^k F/k!. \qquad (122)$$

The equation (122) can be solved in successive approximations in $\gamma$: $F=F_o+\gamma F_1+\gamma F_2/2!+....$ Let's give an example. The expression (104) is rewritten as:

$$\Phi(v_1,v_2,...;X_1,X_2,...)=-(1/2)\sum_{ij}\varphi_{2ij}(v_{ij})X_i(X_j-\delta(i,j));\quad \varphi_{2ij}(v_{ij})=\int_0^\infty(1-exp\{-v_{ij}u_{ij}\})\lambda_{ij}b_{ij}(u_{ij})du_{ij}\ ;(123)$$

$$u_{ij}v_{ij}=u_{i+j}v_{i+j}-u_iv_i-u_jv_j;\quad \lambda_{ij}b_{ij}(u_{ij})=\delta(u_i-1)\delta(u_j-1)\delta(u_{i+j}-1)W(i,j),$$

which coincides with (120) if $\gamma=1$, $r=0$, $k=\delta(k;2)$. Substituting (123) into (103) leads to:

$$\partial F(exp\{-\theta\},t)/\partial t=-(1/2)\sum_{ij}\varphi_{2ij}(\theta_{ij})(\partial^2/\partial\theta_i\partial\theta_j+\delta(i,j)\partial/\partial\theta_i)F(exp\{-\theta\},t), \quad (124)$$

coinciding with (96) if $\theta=-\ln s_i$, $\partial/\partial\theta_i=-s_i\,\partial/\partial s_i$.

The Laplace transform of (124) yields $F(\theta,s)={}_o{}^{\infty}\!\int F(exp\{-\theta\},t)exp\{-st\}dt$,

$$sF(\theta,s)-F(exp\{-\theta\},t=0)=(1/2)\sum_{ij}W(i,j)(exp\{-\theta_{ij}\}-1)[\partial^2/\partial\theta_i\partial\theta_j+\delta(i,j)\partial/\partial\theta_i]F(\theta,s).$$

In [60] the generating functional is written in the Smoluchowsky approximation when $X_i\rightarrow<X_i>$ and $X_j-\delta(i,j)\rightarrow X_j$ in (112) and (123) when the solution for the initial condition $F(exp\{-\theta\},t=0)=exp\{-\sum_i\theta_iX_{oi}\}$ has a form:

$$F_{Sm}(exp\{-\theta\},t)=exp\{-\sum_i\theta_iX_{oi}-\sum_{ij}\varphi_{2ij}(\theta_{ij})\int_0^t<X_i(\tau)><X_j(\tau)>d\tau\}. \qquad (125)$$

The kinetic potential corresponding to the Poisson distribution:

$$\Phi_{Sm}(v,X)=\sum_{ij}[(exp\{v_{ij}\}-1)/2]W(i,j)<X_i><X_j>$$

corresponds to the storage model with the input intensity proportional to $<X_i><X_j>$ and zero release. The expressions for the storage model (104)-(106) correspond to the approximation $\Phi_S(v,X)=\sum_{ij}[(exp\{v_{ij}\}-1)/2]W(i,j)X_{0i}X_{oj}$, when $<X_k>=X_{0k}+(t/2)\sum_{ij}W(i,j)D(i,j|k)X_{0i}X_{0j}$ (D is given in (98)). The time t* of degeneration of the value $N(t)=\sum_k<X_k>/V$, $t^*=2N_o/cM^2$, $W(i,j)=cij$, $M=M_o=\sum_k k<X_k>/V$ coincides with the results of [69] from the Smoluchowski equation. Thus, the model suggested first in Ref. [36] can be generalized at several levels.

The expressions (123), (124) are readily able to yield the equations with higher than (125) precision (involving higher momenta). For example, one gets the refined version of the Smoluchowski equation:

$$\Phi^{(1)}{}_{Sm}(v,X)=\sum_{ij}(exp\{v_{ij}\}-1)W(i,j)X_i<X_j>/2,$$

$$\partial F^{(1)}{}_{Sm}/\partial t=-(1/2)\sum_{ij}W(i,j)(exp\{-\theta_{ij}\}-1)<X_j>\partial F^{(1)}{}_{Sm}(exp\{-\theta\},t)/\partial\theta_i.$$

Rewriting (112) and (123) in the form:

$$\Phi(v,X)=\sum_{ij}(exp\{v_{ij}\}-1)[<X_i><X_j>+\Delta_i(<X_j>-\delta(i,j))-\delta(i,j)<X_i>+\Delta_j<X_i>+\Delta_i\Delta_j]];\quad \Delta_i=X_i-<X_i>,$$

we get:

$-\partial F/\partial\theta_i|_{\theta=0} - <X_i> = <X_i> - <X_i> = 0$, $\partial <X_k>/\partial t = (1/2)\sum_{ij} W(i,j)D(i,j|k)[<X_i><X_j> + \Delta_i\Delta_j - \delta(i,j)<X_i>]$,

two last terms representing the amendments to the Smolukhovsky equation.

From (124) it is possible to derive the "one-particle" Laplace transform of:

$$P_k(X_k) = \int\ldots\int \omega(X_1,\ldots,X_{k-1},X_k,X_{k+1},\ldots)dX_1\ldots dX_{k-1}dX_{k+1}\ldots;$$

$$f_k(\exp\{-\theta_k\}) = \int\ldots\int \exp\{-\theta_k X_k\}\omega(X_1,X_2,\ldots)dX_1dX_2\ldots = \int \exp\{-\theta_k X_k\}P_k(X_k)dX_k.$$

The equation for $f_k$ is:

$$\partial f_k(\exp\{-\theta_k\})/\partial t = \sum_{ij}(1/2)W(i,j)(\exp\{-\theta_k\}-1)D(i,j|k)[<X_iX_j\exp\{-\theta_kX_k\} - \delta(i,j)X_i\exp\{-\theta_kX_k\}>]. \quad (126)$$

With an assertion $<X_iX_k\exp\{-\theta_kX_k\}> \cong -<X_i>\partial f_k/\partial\theta_k$;

$$<X_iX_{k-i}\exp\{-\theta_kX_k\}> \cong <X_iX_{k-i}> f_k \quad (127)$$

yields:

$$(\exp\{-\theta_k\}-1)[\partial^2 f/\partial\theta^2 - a\partial f/\partial\theta - fb] = -(W(k,k))^{-1}\partial f/\partial t; \quad a = (W(k,k))^{-1}[\sum_{i=1,i\neq k}^{\infty} W(k,i)<X_i> - W(k,k)];$$

with following stationary solution:

$$f_{st}(\theta) = [(c-a/2-<X^s_k>)\exp\{(a/2+c)\theta\} + (c+a/2+<X^s_k>)\exp\{(a/2-c)\theta\}]/2c; \quad c = [(a/2)^2+b]^{1/2};$$

$$<(X^s_k)^2> = b - a<X^s_k>, \quad bW(k,k) = \sum_{1\le i\le k-1}(1/2)W(i,k-i)<X_iX_{k-i} - \delta(i,k-i)X_i>.$$

For Poisson distribution:

$$<X^2_k> \cong <X_k>^2 + <X_k>,$$

and $<X_k> = [(a+1)/2]^2 \pm ([(a+1)/2]^2+b)^{1/2}$.

Setting $<X^s_i> = \delta(i,M)$, get $a = w(k,M)/w(k,k)$, $b=0$; $<X^s_k> = w(k,M)/w(k,k) = 1$ by $k=M$.
The approach of this chapter allows the representation of the birth-and-death and some other kinds of processes in the frame of the generalized storage models. New methods of asymptotic analysis for nonlinear perturbed stochastic processes based on new types of asymptotic expansions for perturbed renewal equation and recurrence algorithms for construction of asymptotic expansions for Markov type processes with absorption are presented in Refs. [70-71].

In Ref. [17] a more detailed application of the storage model to the number of monomers in a cluster is also carried out. A more accurate description of the kinetics of aerosol clusters is obtained than using the Smoluchowski equation. Different characteristics of clusters are connected one to another being in fact different aspects of the same problem. For example, the concentration of clusters n(m,t) from (110) is expressed through the number of clusters from (94), (112): $n(m,t) = <X_m>/L^3$. This fact allows us to proceed from the models like (94), (107) to (106)-(107) (and vice versa) making use of the peculiarities of the behavior of aerosols found by means of some specific class of models. For example, the external field yields the additional terms in the right hand of equations of (97), (100). When using the equations like (107) these terms are interpreted as some effective release of monomers (plus complication of the problem due to the spatial inhomogeneity).

One of the advantages of the stochastic storage model is the clear criteria stated for the existence of stationary states in a system Refs. [13, 18]. Besides the external fields, the stationary conditions are determined by such factors as the presence of sources and sinks of clusters, evaporation, condensation, splitting etc.

A generic (typical one) distribution for the storage model is the gamma-distribution (like the Gaussian distribution occupying the same place in the diffusion models). The gamma distribution is known from experiments to fit well the real cluster-size distributions. The storage model is thus open to further detalization of cluster behaviour.

**11c. Stochastic model of domain kinetics in biological membranes**

In Ref. [28] a stochastic model of inventory storage based on the behavior of macroscopic system variables is used to describe the kinetics of raft-type domains in biological membranes.

A lipid membrane is a dynamic ever-changing environment that responds to all the events happening in and around the cells. Certain areas of the membrane are self-organized into cholesterol-rich rafts, clusters of lipid-protein domains, denser and more ordered than other areas and thus drifting freely in the environment, partially isolated regions of the bilayer with specific structural properties. These formations can combine on large platforms, and then the protein molecules, which used to be at different rafts, get an opportunity to meet each other and to interact. In restricted areas of the membrane called lipid rafts, a selective fixation of proteins and lipids is observed. Certain classes of proteins are associated with rafts. Stochasticity, as in the papers [1, 28], is introduced by external noise.

Biological membranes are open thermodynamic systems in a great nonequilibrium state. Many processes occur in them stochastically. In Ref. [28] studies the kinetics of raft-like domains in the membrane under continuous recycling using the stochastic storage model. This task is studied in Ref. [72]. The stochastic approach makes it possible to identify a number of important features of the system behavior that cannot be detected by the deterministic approach.

The processes of formation of domains are reduced to the aggregation (association, clustering) of molecules or ions. The processes of crushing and separation of clusters occur. Such processes are described by the storage models. Apply the exponential distribution to the description of a single jump value. When choosing the exponential distribution $g(x)$ in Eqs. (2), (17) the stationary probability density corresponding to the Laplace transform (11) is equal to (18), (19). Thus, the non-stationary solution for the first moment $< q(t) >$ is Eq. (12).

Parentheses in expressions of the form $< q(t) >$ mean averaging. And dimensionless concentrations $c_n(t)$ in [72] are defined as $c_n(t) = r_n s / S$, where $r_n$ is a number of clusters (domains, rafts) containing $n$ particles, $S$ is a membrane surface area, and $s = \pi (b/2)^2$ is an area of an effective raft monomer with a diameter $b$, $b \sim 5$ nm (1 nm = $10^{-9}$ m). Effective monomer units in Ref. [72] are identified with a protein molecule belonging to a raft together with its associated lipid "skirt". This description is substantially simplified. This issue is discussed below.

The basic assumption made is that the matching of the evolution of raft-like domains in membranes and the storage model (1)-(9) will be adequate. Indeed, monomer molecules and their clusters enter and leave the molecular aggregate (domain), which is the essence of the model (1). The concentration of domains and their number $r_n$ with a certain value of monomers in them is regarded as a variable of the stochastic storage system.

Apply the stochastic storage model Refs. [22, 5, 9, 4] to the description of the behavior of domains in biological membranes examined in Ref. [72]. In Ref. [72] (the equation (3) from Ref. [72]), the kinetic equation is written as a Smoluchowski equation taking into account the summand $\sigma(n)$ describing the recycling of lipids on the cell membrane:

$$\frac{dc_n}{dt} = \sigma(n) + \sum_{m=1}^{\infty} (k_{n,m} c_{n+m} - k`_{n,m} c_n c_m) + \frac{1}{2} \sum_{m=1}^{n-1} (k`_{m,n-m} c_{n-m} c_m - k_{m,n-m} c_n), \qquad (128)$$

where $k_{n,m}$ and $k`_{n,m}$ are rates of domain separation, when one domain of $n+m$ monomers in size is divided into two - $n$ and $m$ in size, and domain consolidation/fusion when two domains of $n$ and $m$ monomers are consolidated into one domain of $n+m$ monomers, $c_n$ is a dimensionless concentration of domains containing $n$ monomers, the domain size distribution function, a random value, the function $\sigma(n)$ describes the phenomenon of the cell membrane recycling and how lipid clusters randomly enter and leave the membrane. In this case, the recycling dynamics is competing with intramembrane kinetics.

First, consider the "scissionless" case of large values of the linear tension $\gamma$ [72], when in the relations obtained in Ref. [72] $k_{n,m} = \exp\{-\gamma[(n)^{1/2} + (m)^{1/2} - (n+m)^{1/2}]\}k`_{n,m}$, at large values $\gamma$ the cluster separation rate $k_{n,m}$ can be neglected. The kinetic equation (128) for the distribution of the domain sizes [72] in this case is simplified and becomes the equation (6) from [72] of the form:

$$dc_n / dt = \sigma(n) - k`Nc_n + k`\sum_{m=1}^{n-1} c_{n-m}c_m / 2, \qquad N = \sum_{m=1}^{\infty} c_m . \tag{129}$$

The paper [72] studies two quite idealized examples (130) and (135) from a broad class of possible membrane recycling schemes (a lot of other schemes can be offered). The first scheme examined is "monomer deposition/raft removal" (MDRR), in which raft proteins and lipids enter the membrane randomly as single-dimension contingent monomer rafts at a rate $j_{on}$, probably in bubbles (bubble transport) belonging to the membrane sections without rafts. Whole rafts leave the membrane at a rate $j_{off}$ independent of their size. The corresponding expression for MDRR mode is Ref. [72]:

$$\sigma(n) = j_{on}\delta_{n,1} - j_{off}c_n , \tag{130}$$

where $\delta_{n,1}$ is Kronecker's delta. In (93)-(103), $k`=1/\tau_D = D/s \approx 10^5$ 1/sec is a domain fusion rate, $\tau_D$ is a characteristic microscopic diffusion time, $D$ is a characteristic diffusion coefficient, at $b \sim 5$ nm, $D \sim 2\times10^{-12}$ m$^2$/c, $\gamma / 2(\pi s)^{1/2}$ is a linear tension, $N = \sum_{m=1}^{\infty} c_m \approx (2 j_{on}\tau_D)^{1/2}$ is a full density of domains of all sizes, $j_{on}$ is a rate of random deposition of single-dimension rafts on the membrane, and $j_{off}$ is an output rate of domains leaving the membrane. These designations are taken from Ref. [72].

Write the equations (129)-(130) as a stochastic equation of the storage theory [4, 5, 9]. To make (129)-(130) correspond to the scheme of the storage model with the stochastic equation of form (1), consider the equation (1) with an output model of form (9). While regarding $c_n$ as a random value, assume that the equations of the form (129)-(130) are true for the random value $c_n$ , which we will consider as a random stock in the system.

In accordance with the assumption made the clusters are formed by jumps. Set the mean value $\mu^{-1}$ of the random inventory entering the system per single jump as $\mu^{-1} = Y_n s / S$, where $Y_n$ is a random number of clusters consisting of $n$ particles entering the ensemble that contain $r_n$ clusters consisting of $n$ monomers per single jump. In the summation $a = k`Nc_n$ in (129), substitute the random concentration $c_n$ by its mean value $\langle c_n \rangle$. The input function is described by relations (2), and the output function – by the expression (9).

The relations (129)-(130) and (1) are matched at:

$$X(t) = c_n; \quad dA(t)/dt = j_{on}\delta_{n,1} + k`\sum_{m=1}^{n-1} c_{n-m}c_m / 2; \quad a = k`N\langle c_n \rangle; \quad b = j_{off} . \tag{131}$$

In the stationary case $\langle dA/dt \rangle = \rho = \lambda / \mu = (a + b\langle c_n \rangle)$:

$$\mu^{-1} = Y_n s / S , \qquad \lambda = \mu(a + b\langle c_n \rangle) = (a + b\langle c_n \rangle)/(Y_n s / S) . \tag{132}$$

The representation of the domain kinetics in the form of a stochastic relation of storage theory allows you to: 1) find the stationary distributions of the random domain concentration value;

2) investigate the noise-induced phase transitions occurring in the system; 3) obtain random concentration moments of any order; 4) write the fluctuation-dissipation relations; and 5) strictly investigate the lifetimes. This also allows obtaining a number of other relations quite important for a more detailed understanding of stochastic processes in biological membranes. Thus, for the stationary distribution of the random value $c_n$, the expressions (18), (19) are written in the model with the output (9) and the input functions (2).

Since Refs. [22, 5, 9, 4] Eq. (26) is satisfied, where $P_0$ is a stationary extinction probability, $\langle \Gamma(x) \rangle$ is an average lifetime of the domain concentration in the stationary state with the initial value $\langle c_n \rangle_{st0} = x$, $\nu(dx) = \lambda\mu \exp\{-\mu x\} dx$ is a random Lévy measure associated with $g(x)$ of the form (17), the average random lifetime of domains of this size in the stationary state averaged out according to initial positions and obtained from (19), (26) is equal to:

$$\int_0^\infty \langle \Gamma(x) \rangle \nu(dx) = (\frac{\mu a}{b})^{-\lambda/b} \frac{\lambda}{b} e^{\mu a/b} \Gamma(\frac{\lambda}{b}; \frac{\mu a}{b}). \quad (133)$$

By plugging (131)-(132) in here, we find the exact expression for the random lifetime of domains in the membrane in the stationary state. The storage models also provide other, more general, possibilities of determining the lifetime Refs. [22, 5]. The distribution (18) changes a stationary form and the noise-induced phase transition occurs at Refs. [22, 4]:

$$\lambda = a\mu + b \qquad \text{or} \qquad b = 0, \qquad j_{off} = 0. \quad (134)$$

In the stationary case $\lambda = \mu(a + b\langle c_n \rangle)$. By plugging in the expression for $\lambda$ from (132), we get that the raft phase, when $\lambda > a\mu + b$ is the right part of Fig. 1, exists at $\langle c_n \rangle \geq \mu^{-1}$, $\langle r_n \rangle \geq Y_n$. I.e. the phase transition corresponding to the situation $\lambda < a\mu + b$ (Fig. 1, left part) occurs when the number of clusters $Y_n$ entering the system of domains consisting of $n$ particles at one impulse (jump-like) entry exceeds the average number of clusters (domains, rafts) $\langle r_n \rangle$ containing $n$ particles already present in the stationary system. We can assume that $Y_n = 1$ (simultaneous formation of two domains of the same size is unlikely), and the raft phase exists almost all the time.

The phase transition in the model (9) at $\lambda = a\mu + b$ occurs as in the Verhulst model Ref. [1], when the type of the distribution with no maximum becomes the distribution with a maximum (see Fig. 1).

The Figure 1 shows that while at $\lambda < a\mu + b$ the distribution maximum corresponds to the zero value of the concentration, at $\lambda = a\mu + b$ the distribution maximum emerges at nonzero value of the concentration, and a group of domains arises from the rare disparate domains. This is a nonequilibrium phase transition induced by external noise. A transition in the system can be induced by keeping the average state of the environment constant, while changing the intensity of fluctuations in the environment. This transition is typical of the multiplicative noise Ref. [3]. Although the nature of the functional dependence of the stationary probability density changes at $\lambda = a\mu + b$, the first two moments of distribution, the mean value and dispersion do not change [1]. At the transition point $D(t)^{1/2} = \langle q(t) \rangle$, where $D(t)$ is dispersion (see below) and $\langle q(t) \rangle$ is a mean value. At $\lambda < a\mu + b$ fluctuations dominate the autocatalytic growth in the raft population, and its extinction remains an unreliable, but still the most expectable outcome. Also note that the description by means of the diffusion approach, in which basic distributions are the Gaussian distributions, is true for large systems. And small systems such as cell membranes and rafts in them are more adequately described by the storage models with the Poisson random noise, in which the basic distributions are gamma distributions as in (18) Ref. [4].

The bimodal distribution will be at the cubic output function or other input functions. The output function of this kind can be obtained taking into account triple collisions of domains, which are not taken into account in Ref. [72]. Clarification of the above description, for example, the consideration of the input function more complex than (17) also allows you to obtain multiple phase transitions.

The values $\langle c_n \rangle$ for the stationary case are defined in [72]. Non-stationary characteristics are also written, for example, a time-dependent expression for the average concentration (12). The solution to the equation for the second moment $\langle q^2(t) \rangle$ obtained from (10) is written as:

$$\langle q^2(t) \rangle = \langle c_n{}^2(t) \rangle = (\langle q_0{}^2 \rangle - \frac{\rho}{\mu b})e^{-2bt} + \frac{\rho}{\mu b} - 2(a-\rho)\int_0^t \langle q(\tau) \rangle e^{-2(t-\tau)b} d\tau .$$

The equation for $P_0(t)$ from $\langle q(t) \rangle$ (12) is also written. The summand with $P_0(t)$ in (12) is smaller than others. You can neglect it by plugging the values of parameters $\lambda, a, b$ from (131), (132) in (12). In one of the modes discussed in Ref. [72] $b = j_{off} = 10$, $\lambda = j_{on} = 1$, $N = 10^{-2,5}$, $a \approx N$, $\rho \approx 10^{-2,5}$, $\langle c_{n=100} \rangle \approx 10^{-5}$. Then in the approximate solution for $\langle c_{n=100}(0) \rangle = \langle c_{n=100} \rangle_{st} = (\rho - a)/b = (\rho / b)/(1 + k`N / b) = 10^{-5}$, $\langle q^2(t) \rangle \approx 10^{-9} + (q_0{}^2 - 10^{-10})e^{-2\times 10t}$, dispersion is $D(t) = \langle q^2(t) \rangle - \langle q(t) \rangle^2$. The average lifetime of the domain (134) for these values of the parameters is equal to 0.27 sec (the value (133) is dimensionless, since the average lifetime on the left side (133) is averaged out according to the measure $\nu(dx)$ with the reverse time dimension; the dimension is obtained by multiplying (133) by the unit quantity of the time dimension $\lambda^{-1} = 1$ sec). Note that at the phase transition point (134) the stationary value of the dispersion is increased (which corresponds to the general theory of phase transitions) and is equal to $2,2 \times 10^{-5}$ instead of $2 \times 10^{-6}$. The stationary state time for the mean value of the concentration is equal to $1/b = 1/j_{off}$ and $1/2b = 1/2j_{off}$ – for the dispersion. These values are less than 0.1 sec and less than the average lifetime of domains. Thus, the stationary is established in the system of membrane domains. But this is a stationary nonequilibrium state of an open system with the exchange of monomers and clusters with the environment and continuous regeneration.

If we assume that the equilibrium state can be established, its thermodynamic characteristics (mean energy, specific heat, etc.) are determined from the equilibrium statistical sum $Q(\beta)$ of the system equal to:

$$P_0^{-1}(\beta_1) = Q(\beta_1) = \frac{a}{b}\int_0^\infty e^{\beta_1 J(u)} du, \quad J(u) = \frac{1}{b}[\int_0^u \frac{\varphi(y/\beta_1)}{y} dy - \frac{au}{\beta_1}], \quad \beta_1 = \frac{\Delta E_0}{k_B T}.$$

In the model (9), this expression takes the form of (19) and takes into account the parameter $\beta^{-1} = k_B T$, where $k_B$ is a Boltzmann constant, $T$ is an absolute temperature, which is necessary for the differentiation to obtain the mean energy, heat capacity, etc., $P_{0\beta}{}^{-1} = 1 + (\frac{\beta\mu a}{b})^{-\frac{\beta\lambda}{b}} \frac{\beta\lambda}{b} e^{\frac{\beta\mu a}{b}} \Gamma(\frac{\beta\lambda}{b}, \frac{\beta\eta a}{b})$; this is the expression (19) written with regard to $\beta^{-1} = k_B T$. The equilibrium membrane systems can be giant synthetic or membrane vesicles. A raft phase in them is observed using the microscope.

In addition to the distribution $g(x)=\mu\exp(-\mu x)$ from (17), other distributions can be used in (131)-(132). Other values are measured in the same manner.

In addition to the mode (130), it is also necessary to examine other modes, for example, an analogue of enzymatic reactions "monomer deposition/monomer removal" (MDMR) of the form (10) Ref. [22], when raft proteins and lipids leave the membrane in single monomers regardless of the size of the raft in which they are located, and:

$$\sigma(n)=j_{on}\delta_{n,1}-j_{off}[nc_n-(n+1)c_{n+1}], \qquad (135)$$

as well as the general equation of the form (3) from Ref. [72] of the form (130). The recycling models (130) and (135) describe the extreme cases of membrane recycling. Other possible schemes fall within these limits.

For storage processes, one can find boundary functionals considered in Ref. [58]. As an example, consider the time that the random variable $X(t)=c_n(t)$ from formulas (129)-(130) spends above some value u. To do this, we should specify the scaled cumulant generating function (SCGF) of the process (1) with the output function (22), which approximates the behavior of the quantity $c_n(t)$ above. For the magnitude of the jumps, we use distribution (17). In Ref. [58] the Lundberg equation k(r)=s is used, where $Ee^{rX(t)}=e^{tk(r)}$, $X(t)=c_n$. As the SCGF function k(r) we choose the function $\ln E[e^{-\theta c_n(t)}|c_n(t=0)]=-\theta c_0e^{-bt}-(\lambda/b)\ln(\mu+\theta e^{-bt})/(\mu+\theta)]$ from expression (24) for the output function of the form (22), as on p. 137, when in expression (137) bw>>a. The same result is written directly from expression (129), when $dc_n/dt=dA(t)/dt-bc_n$, $X(t)=c_n$, $dA(t)/dt=j_{on}\delta_{n,1}+k`\sum_{m=1}^{n-1}c_{n-m}c_m/2;\quad b=j_{off}+k'N$. The parameters included in these functions will be chosen as for the values of the parameters used in (131), (135), and in Ref. [72], when $j_{off}\tau_D=10^{-1}$, $\tau_D=10^{-5}$s, $j_{off}=10$, $N=10^{-2.5}$, $k`=1/\tau_D$, $<c_{n=100}>=10^{-5}$, $\lambda=j_{on}=1$, $\mu=\lambda/b<c_n>$. The Lundberg equation for r takes the form $rc_0e^{-bt}-(\lambda/b)\ln(\mu-re^{-bt})/(\mu-r)]=e^s$. For large t, this equation has a positive solution $\rho_+(s)=\mu(1-e^{-s/(\lambda/b)})$. The same result is obtained from the stationary expression (25), when $k(r)=\ln F(e^r)=-\int_0^{-r}(\varphi(u)/u)du/b=-\lambda\ln(1-r/\mu)/b$. In order to obtain the dependence of $\rho_+(s)$ on $\rho_+(s=0)\neq 0$, we take into account the values of $\varepsilon=e^{-bt}$ at some fixed time in the Lundberg equation, since finite time values are considered. Then at $c_0\approx 10^{-2}$, $t\approx 10^{-2}$, $\varepsilon\approx e^{-1.4}$ $\rho_+(s)\approx b_3(1+c_0\varepsilon\mu e^s/b_3^2)$, $b_3=e^s+c_0\varepsilon\mu-\lambda(1-\varepsilon)/b$.

The behavior of the function $\rho_+(s)$ of s is shown in Fig. 4a.

We use this expression in the relation obtained in [58] for the average values of $Q_u$, the time spent by the random process $X(t)=c_n(t)$ above the value of u, which has the form:

$$\boldsymbol{E}[Q_u,s]=\frac{\boldsymbol{E}[Q_ue^{-sQ_u}]}{\boldsymbol{E}e^{-sQ_u}}=\frac{(\rho_+/(\rho_+(s))^2)(\partial\rho_+(s)/\partial s)e^{-\rho_+u}}{1-(1-\rho_+/\rho_+(s))e^{-\rho_+u}},\ \rho_+=\rho_+(s=0).$$

Calculation results for the dependence $\boldsymbol{E}[Q_u(s)]$ at u=1, Fig.4b, and $\boldsymbol{E}[Q_u(u)]$ at s=0.5, Fig.4c, dependence on the level u, above which the raft size is located.

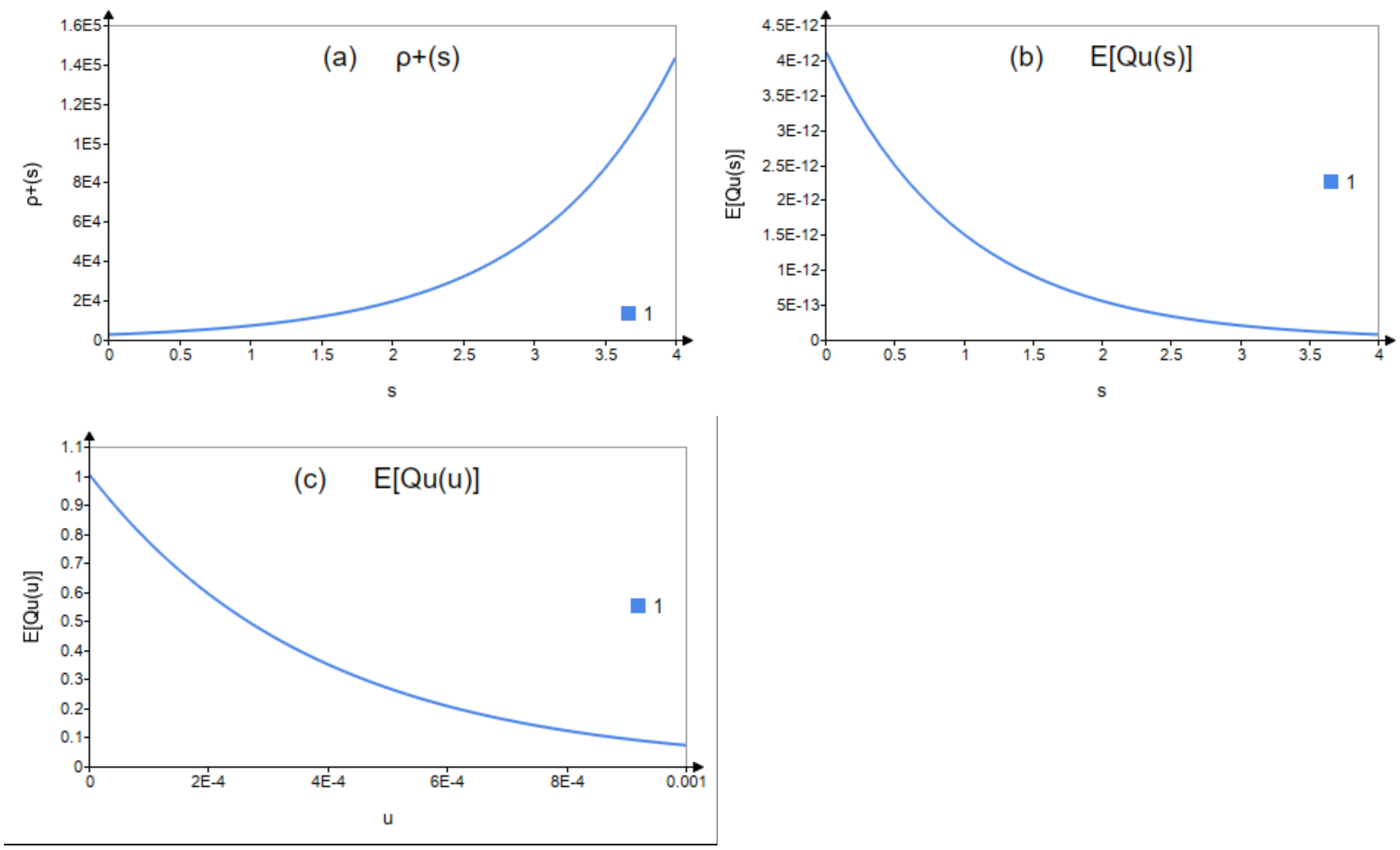

**Fig.4**. a) Behavior of the function ρ+(s) of s;s=0,…4, $c_0 \approx 10^{-2}$, $t \approx 10^{-2}$, b) dependence $\boldsymbol{E}[Q_u(s)]$ at u=0.01, the average time the value $c_n(t)$ stays above the level u=0.01 on the parameter s conjugate to the random variable $Q_u$; c) dependence $\boldsymbol{E}[Q_u(u)]$ at s=0.005, the average time the value $c_n(t)$ stays above the level u=0,0.00001,…,0.001 on the parameter $u=c_n$; $j_{off}\tau_D=10^{-1}$, $\tau_D=10^{-5}$s, $j_{off}=10$, $N=10^{-2.5}$, $k`=1/\tau_D$, $\langle c_{n=100}\rangle=10^{-5}$, $\lambda=j_{on}=1$, $\mu=\lambda/b\langle c_n\rangle$, $b=10+10^{2.5}$.

For the storage models, except for the phase transitions of the form (134), or $\langle r_n\rangle \ge Y_n$ Ref. [28], similar to noise-induced transitions [1] examined in [4] as the behavior of the maximum of the distribution function when random external effects qualitatively change the macroscopic behavior of the system, the conditions for the existence of stationary states are known [9]. Violation of these conditions leads to the restructuring of the system and change in the form of its elements, and from the physical point of view it is also a phase transition. A necessary and sufficient condition for the existence of stationary states is formulated in Ref. [9]: there is a certain value $w_0$, when the equations (38), (41) are satisfied.

After calculating the integral for the case (17), relation (41) takes the following form:

$$\sup_{w\ge w_0}\{-\frac{\lambda}{b}e^{(\frac{a}{b}+w)}Ei[-(\frac{a}{b}+w)]\}<1, \qquad (136)$$

where $Ei(-x)$ is an integral exponent Ref. [40]. By expanding the function $Ei(-x)$ into series at large values of $w$, we find that the existence of stationary states requires the existence of the value $w_0$ that will satisfy the following relation:

$$\sup_{w\ge w_0}\{\frac{\lambda}{b}[\frac{1}{\frac{a}{b}+w}-(\frac{1}{\frac{a}{b}+w})^2+...]\}<1. \qquad (137)$$

At $bw \ll a$, we get the condition for the existence of stationary distributions of the form $\lambda \le a$, $Y_n \ge S/s$ from (137) when limiting the expansion into series by the first summand. However, the condition $bw \ll a$ is satisfied quite rarely. When this condition is satisfied the raft phase will exist in nonstationary states, since the relation $\lambda \leq a,\ Y_n \geq S/s$ is apparently not satisfied. But the output model in the relations (131)-(132) was introduced using artificial means by replacing $a = k`Nc_n$ by $a = k`N\langle c_n \rangle$. It seems that the output model $r(c_n) = bc_n$ is more suitable for rafts. Therefore, in (9) you can assume that $a \to 0$ and consider an opposite case in (137), when $bw >> a$. At $bw >> a$, we get the condition of the form $\lambda \leq bw,\ \ \langle r_n \rangle \leq wY_n$ from (136)-(137), which is satisfied at quite large values of *w*.

This section examines the capabilities of simulating the behavior of raft-like domains using stochastic storage theory. These capabilities are wider than those described in the paper. Thus, we can consider the arbitrary output functions $r(n)$, for example, $\sim n^{2/3}$, or the function $r(q) = bq - cq^2(1-q)$ $(c, b \geq 0)$ corresponding to the non-linear current-voltage characteristic, etc. The general expression for the stationary solution at any $r(n)$ is given in [4]. Using the relations from Ref. [4], for the input (17) we write a form of the output function corresponding to $f(x)$, an arbitrary continuous portion of the distribution function, as $f_{st}(x)$ in (18), $\dfrac{1}{r(x)} = \dfrac{f_{st}(x)e^{\mu x}}{\lambda \int_0^x f_{st}(y)e^{\mu y}dy}$.

The bimodal distributions are also written when choosing more complex input functions that take into account, for example, the entry of both monomers and clusters of any size into the domains and combine the recycling and kinetics processes. We also write fluctuation-dissipation relations that allow getting the connection between $j_{on}$ and $j_{off}$. Instead of the statistical system of molecular aggregates consisting of $n$ molecules with the concentration $c_n$, a single domain can be regarded as a storage system; a more detailed investigation of flows through the potential barrier can be provided, etc. In addition to the average values of the domain concentration, a storage model makes it possible to determine moments of any order of this random value, the lifetime of domains, and the conditions for phase transitions, i.e. the relations between the parameters at which phase transitions, fluctuation-dissipation relations, etc. occur.

In addition to the characteristics of domains examined above, it is possible to define their other parameters. A separate and extensive subject is the capabilities of describing different kinds of influences on the system by the used approach. The condition for the existence of stationary states (38) plays an essential role in the description of physical, chemical and biological systems similar to domain systems. When this condition is violated, some structures are destroyed and replaced by others. The system extinction probability $P_0(t)$ affects the lifetime and other important characteristics of the system. The expression for the average lifetime of membrane domains in the steady state is given above. But to do it, we need to use more complex and cumbersome expressions than (133). The advantage of the proposed method is its generality.

In Ref. [29], the impact of external influences on characteristics of domains in biological membranes is considered. The expressions from Sections 1–9 are used.

In Ref. [30], the dynamic behavior possibilities of raft-like domains in biological membranes are explored. A possible scenario of the behavior of a raft-like domain system oscillating near the phase transition point of the Verhulst transition type, when the form of the stationary distribution for the concentration of domains changes stepwise, has been considered. A stationary state of the system is also possible at the indicated phase transition point, as well as fluctuations in the state of the system between the modes of extinction and survival, if the analogy with the Verhulst model is applied/used. The system behavior is explored in the framework of the stochastic storage model. This model is compared with the Verhulst model of a biological population. Similarities and differences between the models are highlighted. Other features and characteristics of the dynamic behavior and stationary states of the raft-like domain system are considered.

**12. Conclusion**

Stochastic storage processes can be widely applied to solve various physical, chemical and biological problems. Thus, they were used to consider phase transitions induced by external noise [4], various problems of aerosol theory Refs. [11-17], nuclear reactor theory [24-26], micelle theory [27], raft behavior in cell membranes [28-30], tree growth [21], and other problems, for example, general problems of statistical physics [18, 22-23], probabilistic safety [19-20], radiation damage [73]. But these stochastic processes, as well as related processes of risk theory and queue theory, are not very intensively used, in particular, in physics, although they have significant potential for both physical interpretation and heuristic power. Stochastic storage processes are related to risk processes, which were considered in [58]. For example, in [74] these processes were studied as being very similar. The same close connection exists between storage processes and queue processes. Stochastic storage theory has been widely and effectively applied to determining the time of the first reaching of a certain level by a random process. It is possible that this theory will be equally effective in finding other boundary functionals of random processes Ref. [58].

The connection between stochastic storage processes and dynamic systems is also significant. A dynamic system is any system created by man, physical or biological, that changes over time. Such a connection is traced in the work [49]. Thus, the behavior of the simplest storage processes with additive input on intervals between jumps is determined by some dynamic system (processes with deterministic drift Ref. [49]). In Ref. [49], a new definition of storage processes is given, and criteria for the existence of stationary distributions of storage processes are obtained that are different from those formulated in Ref. [7] (of the form (38), (41)-(42)). The results of Ref. [49] generalize the statements proved in work [7] for processes that are solutions of equations (1).

The correspondence between systems of storage theory and dynamic systems established in Ref. [49] makes it possible to apply storage theory, as well as the closely related risk theory (the application of boundary functionals of which to various situations is considered in Ref. [58]) and queue theory, to very broad classes of physical, chemical and biological (as well as other) systems. Difficulties may arise in obtaining explicit exact solutions of the equations written above. But qualitative analysis, various estimates, approximate solutions can be carried out in any case. For example, in Ref. [58] it is shown that the exponential approximation for the distribution of fluctuations differs slightly from the exact solution. In Ref. [75] consider the Erlang A model, or M/M/m+M queue, with Poisson arrivals, exponential service times, and m parallel servers, and the property that waiting customers abandon the queue after an exponential time. The queue length process is in this case a birth–death process, for which we obtain explicit expressions for the Laplace transforms of the time-dependent distribution and the first passage time. These results are also applicable to the M/M/∞ queue, the M/M/m queue, and the M/M/m/m loss model. Having estimated the closeness of solutions of such models to the "exact" solutions, it is possible to use the results of Ref. [58]. Therefore, such approximations may be useful in studying various specific systems. Although, of course, in each case it is necessary to carry out estimates.

This article does not cover all the possibilities of storage processes. Thus, in Ref. [4] the connection between storage processes and the long-range flights and Lévy flights Refs. [76, 77] is indicated, which is not discussed here. Also given are "Gaussian" and "Storage" schemes of the reconstruction of the random process, definitions of the functions V(θ,B) (30) and R(-y,-v) (50)- (51), based on fluctuation-dissipation relations (as in (115)-(124)). These are far from all the examples of application of storage processes.

The author declares that he has no conflicts of interest.

Data Availability Statement: No datasets were generated or analyzed during the current study.